\documentclass[preprint,3p]{elsarticle}

\usepackage{amsmath,amssymb}

\usepackage{xspace}
\usepackage{geometry}

\usepackage[mathscr]{eucal}
\newcommand\La{\mathscr{L}}

\newcommand\FeynArts{{\tt FeynArts}\xspace}
\newcommand\FormCalc{{\tt FormCalc}\xspace}
\newcommand\CalcHep{{\tt CalcHep}\xspace}
\newcommand\CompHep{{\tt CompHep}\xspace}
\newcommand\SARAH{{\tt SARAH}\xspace}

\newcommand\DR{{\overline{\text{DR}}}}

\title{Automatic calculation of supersymmetric Renormalization Group Equations and Loop Corrections}

\author[fs]{Florian Staub} 
\address[fs]{Institut f\"ur Theoretische Physik und Astrophysik, Universit\"at W\"urzburg,\\
D-97074  W\"urzburg, GERMANY}
\ead[fs]{florian.staub@physik.uni-wuerzburg.de}

\begin{document}

\begin{abstract}
\SARAH is a Mathematica package for studying supersymmetric models. It calculates for a given model the masses, tadpole equations and all vertices at tree-level. This information can be used by \SARAH to write model files for \CalcHep/\CompHep or \FeynArts/\FormCalc. In addition, the second version of \SARAH can derive the renormalization group equations for the gauge couplings, parameters of the superpotential and soft-breaking parameters at one- and two-loop level. Furthermore, it calculates the one-loop  self-energies and the one-loop corrections to the tadpoles. \SARAH can handle all \(N=1\) SUSY models whose gauge sector is a direct product of \(SU(N)\) and \(U(1)\) gauge groups. The particle content of the model can be an arbitrary number of chiral superfields transforming as any irreducible representation with respect to the gauge groups. To implement a new model, the user has just to define the gauge sector, the particle, the superpotential and the field rotations to mass eigenstates. 
\end{abstract}

\maketitle

\section*{Program Summary}
{\bf Manuscript Title}: Automatic Calculation of supersymmetric Renormalization Group Equations and Self Energies \\
{\bf Author}: Florian Staub \\
{\bf Program title}: SARAH \\
{\bf Programming language}: Mathematica \\
{\bf Computers for which the program has been designed}: All Mathematica
is available for \\
{\bf Operating systems}: All Mathematica is available for \\
{\bf Keywords}: Supersymmetry, model boodling, Lagrangian, Renormalization Group
Equations, mass matrices, Feynman rules, one loop calculations \\
{\bf CPC Library Classification}: 11.1, 11.6 \\
{\bf Nature of problem}: A supersymmetric model is usually characterized by the particle
content, the gauge sector and the superpotential. It is a time consuming way to obtain
all necessary information for phenomenological studies from this basic ingredients. \\
{\bf Solution method}: SARAH calculates the complete Lagrangian for a given model whose gauge sector can be any direct product of SU(N) gauge groups. The chiral superfields can transform as any, irreducible representation with respect to these gauge groups and it is possible to handle an arbitrary number of symmetry breakings or particle rotations.  Also
the gauge fixing terms can be specified. Using this information, SARAH derives the mass matrices and Feynman rules at tree-level and generates model files for CalcHep/CompHep and FeynArts/FormCalc. In addition, it can calculate the 
renormalization group equations at one- and two-loop level and the one-loop corrections to the one- and two-point functions.  \\
{\bf Unusual features}: SARAH just needs the superpotential and gauge sector as input and not the complete Lagrangian. Therefore, the complete implementation of new models is done in some minutes. \\
{\bf Running time}: Measured CPU time for the evaluation of the MSSM on an Intel Q8200 with
2.33GHz. Calculating the complete Lagrangian: 12 seconds. Calculating all vertices: 75 seconds. Calculating the one- and two-loop RGEs: 50 seconds. Calculating the one-loop corrections: 7 seconds. Writing a FeynArts file: 1 second. Writing a CalcHep/CompHep file: 6 seconds. Writing the LaTeX output: 1 second.

\section{Introduction}
The LHC has started its work and will hopefully find soon the first clues for physics beyond the standard model (SM). Supersymmetry (SUSY) is the most prominent and well studied extension of the SM. However, most studies were done in the context of the Minimal Supersymmetric Standard Model (MSSM) \cite{susy1,susy2,susy3,susy4,susy5,susy6}. Therefore, many computer tools can handle the MSSM  out of the box but it demands some effort to implement new models. Also the analytical expressions for all possible interactions, renormalization group equations or self-energies for the MSSM are nicely presented in literature (see e.g. \cite{Rosiek:1995kg,Kuroda:1999ks,Martin:1993zk,Pierce:1996zz}). Of course, this is not the case for all possible extensions of the MSSM. The Mathematica package \SARAH is supposed to close this gap. The first version \cite{sarah1} already allowed a comprehensive analysis of supersymmetric models by calculating the tree-level masses and tadpole equations as well as all interactions of the model. The obtained information can either be written in \LaTeX{} files or used to generate model files for \CalcHep/\CompHep \cite{CHep}  and \FeynArts/\FormCalc \cite{FeynArts}. To calculate all of  these results, only the minimal amount of information about a model is needed: the gauge sector, the particle content, the superpotential and the field rotations. \\
The new version of \SARAH goes one step further. First, the set of possible models which can be handled has been significantly extended. \SARAH is no longer restricted to chiral superfields in the fundamental representation, but can work with any irreducible representation of \(SU(N)\). Second, \SARAH provides now functions for the calculation of the one-loop masses and the renormalization group equations (RGEs) at one- and two-loop level: it calculates the anomalous dimensions for all chiral superfields and the \(\beta\)-functions for all gauge couplings, superpotential parameters, soft-breaking parameters and vacuum expectation values (VEVs). Furthermore, it calculates the one-loop self-energies of all fields as well as for the one-loop corrections to the tadpoles. This information can be  easily used to get the radiative corrections to the masses at one-loop level. \\
Before we discuss the new features of \SARAH 2, we give a brief, general introduction to \SARAH: in sec.~\ref{sec:down} we explain the installation and the first evaluation of a model. Afterwards, in sec.~\ref{sec:tree}  we show  the main features for obtaining tree-level results and producing model files for \CalcHep/\CompHep or \FeynArts/\FormCalc. In sec.~\ref{sec:rge}, we discuss the possibilities to derive one- and two-loop RGEs, before we show in sec.~\ref{sec:loop} how to calculate the one-loop corrections to one- and two-point functions. Finally, we explain how to implement new models in \SARAH in sec.~\ref{sec:model}.\\
The appendix contains supplementary information about the check for gauge anomalies (\ref{sec:GaugeAnomaly}), the calculation of the Lagrangian in component fields (\ref{sec:Lagrangian}) and the derivation of gauge group factors (\ref{sec:irrep}). In addition, we list all formulas used for the calculation of the RGEs and self-energies in \ref{app:formulas}, show our conventions for the MSSM in \ref{sec:MSSM} and \ref{sec:SARAH_MSSM}, before we summarize the changes in comparison to the first version of \SARAH in \ref{app:changes}.

\section{Download, installation and first evaluation}
\label{sec:down}
\SARAH can be downloaded from
\begin{verbatim}
http://theorie.physik.uni-wuerzburg.de/~fnstaub/sarah.html
\end{verbatim}
The package should be extracted to the application directory of Mathematica. This directory under Linux in
\begin{verbatim}
home/user/.Mathematica/Applications/
\end{verbatim}
and
\begin{verbatim}
Mathematica-Directory\AddOns\Applications\
\end{verbatim}
in Windows. \\
Initially, the package itself consists  of two directories: the directory \verb"Package" contains all Mathematica package files, while in the directory \verb"Models" the definitions of the different models are located. During the work, a third directory called \verb"Output" is generated by \SARAH. The results of different calculations as well as the model files for the diagram calculators are written to this directory. A fourth directory \verb"LaTeX" contains \LaTeX{} packages which are needed for the appropriate output. \\
A comprehensive manual ({\tt sarah.pdf}) is included in the package archive and can also be found on the web page and on arXiv \cite{sarah}. In addition, a file ({\tt models.pdf}) with information about all models delivered with the package is part of the archive. Furthermore, a file with a short introduction to the main commands is included ({\tt Readme.txt}) as well as an example for the usage ({\tt Example.nb}).  \\
After the installation, the package is loaded in Mathematica via
\begin{verbatim}
<<"sarah-2.0/SARAH.m"
\end{verbatim} 
and a supersymmetric model is initialized by 
\begin{verbatim}
Start["Modelname"];
\end{verbatim} 
Here, \verb"Modelname" is the name of the corresponding model file, e.g. for the minimal supersymmetric standard model the command would read
\begin{verbatim}
Start["MSSM"];
\end{verbatim} 
or for the next-to-minimal supersymmetric model in CKM basis
\begin{verbatim}
Start["NMSSM","CKM"];
\end{verbatim}
is used. In the following, we refer for all given examples the model file of the MSSM. Our conventions concerning the fields definitions and rotations in the MSSM are given in \ref{sec:MSSM}. Even if the meaning of most symbols used in the examples should be intuitively clear, we list the internal names for parameters and particles in \ref{sec:SARAH_MSSM}. In addition, we give in \ref{app:changes} an overview about the main changes happened in the second version of \SARAH in comparison to the first version presented in \cite{sarah1}.
\section{Tree-level calculations}
\label{sec:tree}
When a model is initialized using the {\tt Start} command, this model is first checked for gauge anomalies and charge conservation. If not all checks are fulfilled, a warning is printed. More information about the different checks is given in \ref{sec:GaugeAnomaly}. Afterwards, the calculation of the complete Lagrangian at tree-level starts. The performed steps are presented in \ref{sec:Lagrangian}. When this calculation is finished, several tree-level results can easily be obtained.
\subsection{Particle content}
To get an overview of all particles of the different eigenstates, 
\begin{verbatim}
Particles[Eigenstates]
\end{verbatim}
is used. This can be, for instance, \verb"Particles[GaugeES]" or \verb"Particles[EWSB]" for the gauge eigenstates or the eigenstates after electroweak symmetry breaking (EWSB), respectively. The output is a list with the following information about each particle: (i) name of the particle, (ii) type of the particle (\verb"F" for fermion, \verb"S" for scalar, \verb"V" for vector boson, \verb"G" for ghosts, \verb"A" for auxiliary field), (iii) number of first generation, (iv) number of last generation, (v) indices of the particle. Fermions are listed using Weyl and not Dirac spinors. For instance, the entry for the gauge eigenstates of the left-down quark reads
\begin{verbatim}
{FdL, 1, 3, F, {generation, color}}
\end{verbatim}
\subsection{Masses and tadpole equations}
The masses and tadpole equations are derived automatically during the evaluation of a model. In this regard, the masses or the entries of a mass matrix are calculated as second derivative of the Lagrangian
\begin{equation}
 m_{i j} = - \frac{\partial^2 \La}{\partial \phi_i \partial \phi_j^*}
\end{equation}
with respect to the considered fields \(\phi_{i}\). The tadpoles \(T_i\) are the first derivative of the scalar potential with respect to the different VEVs \(v_i\)
\begin{equation}
\frac{\partial V}{\partial v_i} \equiv T_i \thickspace .
\end{equation}
The user has access to both information by using the command {\tt MassMatrix[Particle]} for the mass matrix of  \verb"Particle" and 
{\tt  TadpoleEquation[VEV]} for the tadpole equation corresponding to the vacuum expectation value \verb"VEV".
\paragraph*{Examples}
\begin{enumerate}
\item {\bf Higgs mass matrix}. The \((1,2)\)-entry of the mass matrix of the scalar Higgs in the MSSM is saved in
{\tt MassMatrix[hh][[1,2]]}. This returns
\begin{verbatim}
-(g1^2*vd*vu)/4 - (g2^2*vd*vu)/4 - B[\[Mu]]/2 -conj[B[\[Mu]]]/2 
\end{verbatim}
\item {\bf Squark mass matrix}. In the same way, the (1,1)-entry of the \(6 \times 6\) down squark mass matrix is obtained by
 {\tt MassMatrix[Sd][[1,1]]}. The output is
\begin{verbatim}
(-3*g2^2*(vd^2 + vu^2) + g1^2*(vu^2 - vd^2) + 24*mq2[1,1] +
    12*vd^2*sum[j1, 1, 3,conj[Yd[j1, 1]]*Yd[j1, 1]])/24
\end{verbatim}
\item {\bf Squark mass matrix with generation indices as variable}. To get the result for the \(2 \times 2\) down squark matrix without the explicit insertion of generation indices, {\tt MassMatrixUnexpanded[Sd][[1,1]]} is used. The output is
\begin{verbatim}
(Delta[cm1,cn1]*(-((g1^2+3*g2^2)*(vd^2-vu^2)*Delta[gm1,gn1]) 
  + vd^2*sum[j1,1,3,conj[Yd[j1,gn1]]*Yd[j1,gm1]] + 12*(2*mq2[gm1,gn1] )))/24
\end{verbatim}
\item {\bf Tadpole equation}. The tadpole equation corresponding to  \(\frac{\partial V}{\partial v_d}=0\) is obtained by 
{\tt TadpoleEquation[vd]  } and reads
\begin{verbatim}
(8*mHd2*vd + g1^2*vd^3 + g2^2*vd^3 - g1^2*vd*vu^2  - g2^2*vd*vu^2 - 4*vu*B[\[Mu]] +
( 8*vd*\[Mu]*conj[\[Mu]] -  4*vu*conj[B[\[Mu]]])/8 == 0
\end{verbatim}
\end{enumerate}
\subsection{Vertices}
The vertices are calculated as partial derivatives of the Lagrangian with respect to the involved fields and applying afterwards the vacuum conditions. The vertices can be calculated in two ways. Either it is possible to calculate the vertices for a specific choice of external particles or to calculate all vertices of the complete model at once. The former task is evolved by
\begin{verbatim}
 Vertex[{Particles},Options];
\end{verbatim}
The argument of this function is a list with the external particles. The options define the set of eigenstates ({\tt Eigenstates}\ \(\rightarrow\) name) and usage of relations among the parameters ({\tt UseDependences \(\rightarrow\) True/False}). In the result, the coefficients corresponding to different Lorentz structures are separately listed.  If possible, the expressions are simplified by using the unitarity of rotation matrices, the properties of generators and, if defined, simplifying assumptions about parameters.\\
All vertices for a set of eigenstates are calculated at once  by
\begin{verbatim}
MakeVertexList[Eigenstates, Options];
\end{verbatim}
This searches for all possible interactions present in the Lagrangian and creates lists for the generic subclasses of interactions, e.g. {\tt VertexList[FFS]} or {\tt VertexList[SSVV]} for all two-fermion-one-scalar interactions and all two-scalar-two-vector-boson interactions, respectively. If effective theories are considered, six-particle interaction ({\tt SixParticleInteractions \(\rightarrow\) False}) or all higher-dimensional operators ({\tt effectiveOperators \(\rightarrow\) False}) can be switched off during this calculation. Those interactions slow down the computation and they are sometimes not needed. 
\paragraph*{Examples}
\begin{enumerate}
\item {\bf One possible Lorentz structure}. {\tt Vertex[\{hh,Ah,Z\}] } leads to the vertex of scalar and a pseudo scalar Higgs with a Z-boson
\begin{verbatim}
{{hh[{gt1}], Ah[{gt2}], VZ[{lt3}]}, 
 {((ZA[gt2,1]*ZH[gt1,1] - ZA[gt2,2]*ZH[gt1,2])*(g2*Cos[ThetaW]+g1*Sin[ThetaW]))/2, 
                Mom[Ah[{gt2}], lt3] - Mom[hh[{gt1}],lt3]}}
\end{verbatim}
The output is divided in two parts. First, the involved particles are given, second, the value of the vertex is given. This second part is again split in two parts: one is the Lorentz independent part and the second part defines the transformation under the Lorentz group.  
\item {\bf Several possible Lorentz structures}. {\tt Vertex[\{bar[Fd],Fd,hh\}]} is the interaction between two  d-quarks and a Higgs:
\begin{verbatim}
{{bar[Fd[{gt1, ct1}]], Fd[{gt2, ct2}], hh[{gt3}]}, 
 {((-I)*Delta[ct1,ct2]*Delta[gt1,gt2]*ZH[gt3,2]*Yd[gt2,gt1])/Sqrt[2],PL}, 
 {((-I)*Delta[ct1,ct2]*Delta[gt1,gt2]*ZH[gt3,2]*Yd[gt1,gt2])/Sqrt[2],PR}}
\end{verbatim}
Obviously, there are three parts: one for the involved particles and two for the different Lorentz structures. \verb"PL" and \verb"PR" are the polarization projectors \(P_L = \frac{1}{2} (1 - \gamma_5), P_R = \frac{1}{2} (1 + \gamma_5)\).
\item {\bf Changing the considered eigenstates and using Weyl fermions} It is also possible to calculate the vertices for Weyl fermions and/or to consider the gauge eigenstates. For instance,
\begin{verbatim}
Vertex[{fB, FdL, conj[SdL]}, Eigenstates -> GaugeES]
\end{verbatim}
gives
\begin{verbatim}
{{fB, FdL[{gt2, ct2}], conj[SdL[{gt3, ct3}]]}, 
 {((-I/3)*g1*Delta[ct2, ct3]*Delta[gt2, gt3])/Sqrt[2],1}}
\end{verbatim}
\item {\bf Using dependences}. With {\tt Vertex[\{conj[Se], Se, VP\}, UseDependences -> True]} \(g_1\) and \(g_2\) are replaced by the electric charge \(e\). This and similar relations can be defined in the parameter file (see sec.~\ref{sec:ParametersFile}). 
\begin{verbatim}
{{conj[Se[{gt1}]], Se[{gt2}], VP[{lt3}]}, 
{(-I)*e*Delta[gt1,gt2],-Mom[conj[Se[{gt1}]],lt3]+Mom[Se[{gt2}],lt3]}}
\end{verbatim}
\item {\bf Fixing the generations}. It is possible to give the indices of the particles already as input
\begin{verbatim}
Vertex[{hh[{1}], hh[{1}], Ah[{2}], Ah[{2}]}]
\end{verbatim} 
leads to
\begin{verbatim}
{{hh[{1}], hh[{1}], Ah[{2}], Ah[{2}]}, 
 {(-I/4)*(g1^2 + g2^2)*Cos[2*\[Alpha]]*Cos[2*\[Beta]], 1}}
\end{verbatim}
Obviously, the given definition of the mixing matrices for the Higgs fields was automatically inserted. 
\end{enumerate}
\subsection{Output for diagram calculators and  \LaTeX}
\paragraph*{\CalcHep/\CompHep} \CalcHep and \CompHep are well known  and widely used programs for calculating cross sections and decay widths via a diagrammatic approach at tree-level. The model files produced by \SARAH support both Feynman gauge and unitarity gauge. Furthermore, \SARAH can split  interactions between four colored particles in a way that they can be handled by \CalcHep/\CompHep and also models with CP violation are possible. The model files for \CalcHep/\CompHep are created by 
\begin{verbatim}
 MakeCHep[Options];
\end{verbatim}
The options define, whether the Feynman gauge should be included ({\tt FeynmanGauge \(\rightarrow\) True/False}) and if CP violation should be considered ({\tt CPViolation \(\rightarrow\) True}). Also the splitting of specific four-scalar interactions can be suppressed as long as they are not colored ({\tt NoSplitting} \(\rightarrow\) list of fields). In addition, the running of the strong coupling constant can be included as it is usually  done in the standard \CalcHep files ({\tt UseRunningCoupling \(\rightarrow\) True}).   
\paragraph*{\FeynArts/\FormCalc} \FeynArts is a Mathematica package for creating Feynman diagrams and the corresponding amplitudes. This information is afterwards used by \FormCalc to simplify the amplitudes and square them by using {\tt FORM}. In contrast to \CalcHep/\CompHep{} \FeynArts/\FormCalc can deal also with loop diagrams. Beside the standard model file for \FeynArts, \SARAH writes a second file including supplementary information about the model: all defined dependences, the numerical values for the parameters and masses, if they are available, and special abbreviations to speed up the calculations with \FormCalc. The model files are generated via
\begin{verbatim}
 MakeFeynArts;
\end{verbatim}
\paragraph*{\LaTeX}
The generated \LaTeX\xspace files include all information about a model for one set of eigenstates: particle content, mixing matrices, tadpole equations, RGEs, one-loop self-energies and one-loop corrections to the tadpoles as well as all interactions. The \LaTeX\xspace output using the standard functions of Mathematica is not really readable when dealing with long expressions. Therefore, special functions were developed for a better typesetting.  For each vertex, the corresponding Feynman diagram  is automatically drawn using \verb"FeynMF" \cite{Ohl:1995kr}. In addition, a batch file for Linux and for Windows are written by \SARAH to simplify the compilation of the different \LaTeX{} files including all Feynman diagrams. The command for producing the \LaTeX{} output is 
\begin{verbatim}
 MakeTeX[Options]; 
\end{verbatim}
One option is to disable the output of the Feynman diagrams ({\tt FeynmanDiagrams \(\rightarrow\) False}), another to use a shorter notation for the interactions ({\tt ShortForm \(\rightarrow\) True}).
\section{Renormalization Group Equations}
\label{sec:rge}
\label{RGEs}
SARAH calculates the RGEs for the parameters of the superpotential, the soft-breaking terms, the gauge couplings at one- and two-loop level and the VEVs. This is done by using the generic formulas of \cite{Martin:1993zk,Yamada:1993ga,Yamada:1993uh, Yamada:1994id} which we have summarized in \ref{app:RGE_con}. \\
The calculation of the RGEs can be started after the initialization of a model via
\begin{verbatim}
CalcRGEs[Options];
\end{verbatim}
\paragraph*{Options}
The options offer a possibility to disable the calculation of the two-loop RGEs (\verb"TwoLoop" \(\rightarrow\) \verb"False"). Another option is to handle the number of generations of specific chiral superfields as variable ({\tt VariableGenerations} \(\rightarrow\) list of fields). This might be used for models which include chiral superfields much heavier than the SUSY scale to make the dependence on these fields obvious. Normally, the \(\beta\)-function are written in a compact form using matrix multiplication, as explained below. This can be switched off by the option \verb"NoMatrixMultiplication"  \(\rightarrow\) \verb"True". The last option offers the possibility to read the results of former calculations (\verb"ReadLists"  \(\rightarrow\) \verb"True")\\
\paragraph*{GUT normalization}
The gauge couplings of abelian  gauge groups are often normalized at the GUT scale. Therefore, it is possible to define for each  \(U(1)\) gauge coupling the GUT-normalization by the corresponding entry in the parameters file. See sec.~\ref{sec:ParametersFile} for more information. 
\paragraph*{Results}
The RGEs are saved in different arrays in Mathematica whose names are given in brackets: anomalous dimensions of all superfields (\verb"Gij"), trilinear superpotential parameters (\verb"BetaYijk"), bilinear superpotential parameters (\verb"BetaMuij"), linear superpotential parameters (\verb"BetaLi"), trilinear soft-breaking parameters (\verb"BetaTijk"), bilinear soft-breaking parameters (\verb"BetaBij"), linear soft-breaking parameters (\verb"BetaLSi"), scalar soft-breaking masses (\verb"Betam2ij"), gaugino soft-breaking masses (\verb"BetaMi"), gauge couplings (\verb"BetaGauge") and VEVs (\verb"BetaVEVs").\\
All entries of these arrays are three-dimensional: the first entry gives the name of the parameter, the second one the one-loop \(\beta\)-function and the third one the two-loop \(\beta\)-function. Furthermore, the results for the RGEs of the scalar masses are simplified by using abbreviations for often appearing traces (see also \cite{Martin:1993zk}). The definition of the traces is saved in the array \verb"TraceAbbr". In \verb"TraceAbbr[[1]]" all abbreviations needed for the one-loop RGEs are given, and in \verb"TraceAbbr[[2]]" those are for the two-loop.\\
The results are also saved as text files in the directory
\begin{verbatim}
PackageDirectory/Output/Modelname/RGEs/ 
\end{verbatim}

\paragraph*{Matrix Multiplication}
Generally, the results contain sums over the generation indices of the particles in the loop. \SARAH always tries to write them as matrix multiplications, in order to shorten the expressions. Therefore, new symbols are introduced:
\begin{itemize}
\item \verb"MatMul[A,B,C,...][i,j]": \((A B C \dots)_{i,j}\). Matrix multiplication, also used for vector-matrix and vector-vector multiplication.
\item \verb"trace[A,B,C,...]": \(\mbox{Tr}(A B C \dots)\). Trace of a matrix or product of matrices.
\item \verb"Adj[M]": \(M^\dagger\). Adjoint of a matrix.
\item \verb"Tp[M]": \(M^T\). Transposed of a matrix.
\end{itemize} 
As already mentioned, the usage of matrix multiplication can be switched off with the corresponding option. In addition, it is automatically switched off, if the model contains a parameter with three generation indices. 
\paragraph*{Examples}
\begin{enumerate}
\item {\bf \(\beta\)-function of Yukawa couplings}. The Yukawa couplings of the MSSM are saved in \verb"BetaYijk". The first entry consists of
\begin{verbatim}
BetaYijk[[1,1]]:  Ye[i1,i2] ,
\end{verbatim}
i.e. this entry contains the \(\beta\)-functions for the electron Yukawa coupling. The corresponding one-loop \(\beta\)-function is
\begin{verbatim}
BetaYijk[[1,2]]:
(-9*g1^2*Ye[i1,i2])/5-3*g2^2*Ye[i1,i2]+3*trace[Yd,Adj[Yd]]*Ye[i1,i2]+ 
  trace[Ye,Adj[Ye]]*Ye[i1, i2]+3*MatMul[Ye,Adj[Ye],Ye][i1, i2]
\end{verbatim}
The two-loop \(\beta\)-function is saved in \verb"BetaYijk[[1,3]]" but we skip it here because of its length. 
\item {\bf \(\beta\)-function of soft-breaking masses and abbreviations for traces}.  The first entry of \verb"Betam2ij" corresponds to the soft-breaking mass of the selectron
\begin{verbatim}
 Betam2ij[[1,1]]:           me2[i1,i2]
\end{verbatim}
 and the one-loop \(\beta\)-function is saved in  \verb"Betam2ij[[1,2]]":
\begin{verbatim}
(-24*g1^2*MassB*conj[MassB]+10*g1^2*Tr1[1])*Kronecker[i1,i2]/5 + 
 4*mHd2*MatMul[Ye,Adj[Ye]][i1,i2]+4*MatMul[T[Ye],Adj[T[Ye]]][i1,i2] + 
  2*MatMul[me2,Ye,Adj[Ye]][i1,i2]+4*MatMul[Ye, ml2, Adj[Ye]][i1,i2] + 
  2*MatMul[Ye,Adj[Ye],me2][i1,i2]
\end{verbatim}
The definition of the element \verb"Tr1[1]" is saved in \verb"TraceAbbr[[1,1]]":
\begin{verbatim}
{Tr1[1], -mHd2 + mHu2 + trace[md2] + trace[me2] - trace[ml2] +
          trace[mq2] - 2*trace[mu2]}
\end{verbatim}
\item {\bf Number of generations as variable}. With
\begin{verbatim}
CalcRGEs[VariableGenerations -> {q}]
\end{verbatim}
the number of generations of the left-quark superfield is handled as variable. Therefore, the one-loop \(\beta\)-function of the hypercharge couplings reads 
\begin{verbatim}
 (63*g1^3)/10 + (g1^3*NumberGenerations[q])/10
\end{verbatim}
\item{\bf No matrix multiplication}. Using matrix multiplication can be switched off by
\begin{verbatim}
CalcRGEs[NoMatrixMultiplication -> True]
\end{verbatim}
The one-loop \(\beta\)-function for the electron Yukawa coupling is now written as
\begin{verbatim}
  sum[j2,1,3,sum[j1,1,3,conj[Yd[j2,j1]]*Yu[i1,j1]]*Yd[j2,i2]] + 
2*sum[j2,1,3,sum[j1,1,3,conj[Yu[j1,j2]]*Yu[j1,i2]]*Yu[i1,j2]] + 
  sum[j2,1,3,sum[j1,1,3,conj[Yu[j2,j1]]*Yu[i1,j1]]*Yu[j2,i2]] + 
(3*sum[j2,1,3,sum[j1,1,3,conj[Yu[j1,j2]]*Yu[j1,j2]]]*Yu[i1,i2])/2 + 
(3*sum[j2,1,3,sum[j1,1,3,conj[Yu[j2,j1]]*Yu[j2,j1]]]*Yu[i1,i2])/2 - 
 (13*g1^2*Yu[i1,i2])/15-3*g2^2*Yu[i1,i2]-(16*g3^2*Yu[i1,i2])/3
\end{verbatim}
\end{enumerate}

\section{Loop Corrections}
\label{sec:loop}
\SARAH calculates the analytical expressions for the one-loop corrections to the tadpoles and the self energy of all particles. These calculations are performed in \(\overline{\mbox{DR}}\)-scheme and in the 't Hooft gauge. In principle, this is a generalization of the calculations for the MSSM presented in \cite{Pierce:1996zz}. The command to start the calculation is
\begin{verbatim}
CalcLoopCorrections[Eigenstates];
\end{verbatim}
As usual, \verb"Eigenstates" can in the case of the MSSM either be \verb"GaugeES" for the gauge eigenstates or \verb"EWSB" for the eigenstates after EWSB. If the vertices for the given set of eigenstates were not calculated before, this is done before the calculation of the loop contributions begins. \\
\paragraph*{Conventions} Using the conventions of \cite{Pierce:1996zz}, the results will contain the  Passarino Veltman integrals listed in \ref{sec:Integrals}. The involved couplings are abbreviated by
\begin{itemize}
\item \verb"Cp[p1,p2,p3]" and \verb"Cp[p1,p2,p3,p4]" for non-chiral, three- and four-point interactions involving the particles \verb"p1" - \verb"p4".
\item \verb"Cp[p1,p2,p3][PL]" and \verb"Cp[p1,p2,p3][PR]" for chiral, three-point interactions involving the fields \verb"p1" - \verb"p3". 
\end{itemize}
The self-energies can be used for calculating the radiative corrections to masses and mass matrices, respectively. We have summarized the needed formulas for this purpose in \ref{sec:OneLoopMass}. For calculating the loop corrections to a mass matrix, it is convenient to use unrotated, external fields, while the fields in the loop are rotated. Therefore, \SARAH adds to the symbols of the external particle in the interaction an \verb"U" for 'unrotated', e.g. \verb"Sd" \(\rightarrow\) \verb"USd". The mixing matrix associated to this field in the vertex has to be replaced by the identity matrix when calculating the correction to the mass matrix. 
\paragraph*{Results} The results for the loop corrections are saved in two different ways. First as list containing the different loop contribution for each particle. Every entry includes the following information: internal particles, generic type of the diagram, numerical factors coming from symmetry considerations and possible charges in the loop. The second output is a sum of all contributions, where the generic results of \ref{sec:Integrals} have already been inserted. This can afterwards be written as pdf file using the \LaTeX{} output of \SARAH.\\
The results for the self-energies are saved in \verb"SelfEnergy1LoopList" as list of the contributions and in \verb"SelfEnergy1LoopSum" written as sums. The last one is a two-dimensional array. The first column gives the name of the external particle, the entry in the second column depends on the type of the field. For scalars the one-loop self energy \(\Pi(p^2)\)  is given,  for fermions the one-loop self-energies for the different polarizations (\(\Sigma^L(p^2)\),\(\Sigma^R(p^2)\), \(\Sigma^S(p^2)\)) are written in a three-dimensional array, while for vector bosons the transversal part of the self energy \(\Pi^T(p^2)\) is shown. \\
Also the results for corrections to the tadpoles are saved twice: in \verb"Tadpoles1LoopSum[Eigenstates]" explicitly written as sum and in \verb"Tadpoles1LoopList[Eigenstates]"  as list of the different contributions.   
\paragraph*{Examples}
\begin{enumerate}
\item {\bf One-loop tadpoles}. The radiative correction of the tadpoles due to a chargino loop is saved in 
\begin{verbatim}
Tadpoles1LoopList[EWSB][[1]]; 
\end{verbatim}
and reads
\begin{verbatim}
 {bar[Cha],Cp[Uhh[{gO1}],bar[Cha[{gI1}]],Cha[{gI1}]],FFS,1,1/2}
\end{verbatim}
The meaning of the different entries is: (i) a chargino (\verb"Cha") is in the loop, (ii) the vertex with an external, unrotated Higgs (\verb"Uhh") with generation index \verb"gO1" and two charginos with index \verb"gI1" is needed, (iii) the generic type of the diagram is \verb"FFS", (iv) the charge factor is 1, (v) the diagram is weighted by a factor \(\frac{1}{2}\) with respect to the generic expression (see \ref{sec:Integrals}).\\
The corresponding term in \verb"Tadpoles1LoopSum[EWSB]" is
\begin{verbatim}
4*sum[gI1,1,2, A0[Mass[bar[Cha[{gI1}]]]^2]*
   Cp[phid,bar[Cha[{gI1}]],Cha[{gI1}]]*Mass[Cha[{gI1}]]] 
\end{verbatim}
\item {\bf One-loop self-energies}.
\begin{enumerate}
\item The correction to the down squark matrix due to  a four-point interaction with a pseudo scalar Higgs is saved in
{\tt  SelfEnergy1LoopList[EWSB][[1, 12]]} and reads
\begin{verbatim}
{Ah,Cp[conj[USd[{gO1}]],USd[{gO2}],Ah[{gI1}],Ah[{gI1}]],SSSS,1,1/2}
\end{verbatim}
This has the same meaning as the term
\begin{verbatim}
-sum[gI1,1,2,A0[Mass[Ah[{gI1}]]^2]*
   Cp[conj[USd[{gO1}]],USd[{gO2}],Ah[{gI1}],Ah[{gI1}]]]/2
\end{verbatim}
in \verb"SelfEnergy1LoopSum[EWSB]". 
\item Corrections to the Z boson are saved in {\tt SelfEnergy1LoopList[EWSB][[15]]}. An arbitrary entry looks like
\begin{verbatim}
{bar[Fd], Fd, Cp[VZ, bar[Fd[{gI1}]], Fd[{gI2}]], FFV, 3, 1/2}
\end{verbatim}
and corresponds to
\begin{verbatim}
(3*sum[gI1, 1, 3, sum[gI2, 1, 3, 
     H0[p^2, Mass[bar[Fd[{gI1}]]]^2, Mass[Fd[{gI2}]]^2]*
 (conj[Cp[VZ,bar[Fd[{gI1}]],Fd[{gI2}]][PL]]*
     Cp[VZ,bar[Fd[{gI1}]],Fd[{gI2}]][PL] + 
  conj[Cp[VZ,bar[Fd[{gI1}]],Fd[{gI2}]][PR]]*
     Cp[VZ,bar[Fd[{gI1}]],Fd[{gI2}]][PR]) + 
 2*B0[p^2,Mass[bar[Fd[{gI1}]]]^2,Mass[Fd[{gI2}]]^2]*
     Mass[bar[Fd[{gI1}]]]*Mass[Fd[{gI2}]]*
  Re[Cp[VZ,bar[Fd[{gI1}]],Fd[{gI2}]][PL]*
     Cp[VZ,bar[Fd[{gI1}]],Fd[{gI2}]][PR])]])/2 
\end{verbatim}
in \verb"SelfEnergy1LoopListSum[EWSB]". 
\end{enumerate}
\end{enumerate}
\section{Definition of models}
\label{sec:model}
All information of the different models is saved in three different files, which have to be in one directory
\begin{verbatim}
PackageDirectory/Models/ModelName/
\end{verbatim} 
The three files are: one model file with the same name as the directory (\verb"ModelName.m"), a file containing additional information about the particles of the model (\verb"particles.m") and a file containing additional information about the parameters of the model (\verb"parameters.m"). Only the first file is really necessary for calculating the Lagrangian and to get a first impression of a model. However,  the other two files are needed for defining properties of parameters and particles and for producing an appropriate output. \\
In addition, it is possible to include LesHouches spectrum files in the model directory \cite{Allanach:2008qq}. These can be read with the command {\tt ReadSpectrum["Filename"]}. If spectrum files are located in another directory, the complete path has to be added to the filename to read them. 

\subsection{The model file}
The model file contains the following parts: first, the gauge structure and the particle content are given. Afterwards, the matter interactions are defined by the superpotential. These are general information which must always be apparent. \\ 
This part is followed by the definition of the names for all eigenstates ({\tt NameOfStates}). For these eigenstates, several properties can be defined using the corresponding {\tt DEFINITION} statement: decomposition of complex scalars in scalar, pseudo scalar and VEV ({\tt DEFINITION[States][VEVs]}), rotations in the matter ({\tt DEFINITION[States][MatterSector]}) and gauge sector ({\tt DEFINITION[States][GaugeSector]}), the corresponding gauge fixing terms ({\tt DEFINITION[States][GaugeFixing]}), the flavor decomposition of fields ({\tt DEFINITION[States][Flavors]}) and possible phases of fields ({\tt DEFINITION[States][Phases]}). New couplings can be added and existing couplings can be changed by hand ({\tt DEFINITION[States][Additional]}).  \\
Afterwards, optionally the particles are states, which should be integrated out or deleted. At the end, the Dirac spinors have to be built out of Weyl spinors, a spectrum file can be defined and a choice for an automatic output can be made. 
\subsubsection{Vector and chiral superfields}
\SARAH supports all \(SU(N)\) gauge groups. The gauge sector in \SARAH is defined by adding a vector superfield for each gauge group to the list {\tt Gauge} in the model file, e.g. 
\begin{verbatim}
Gauge[[3]]={G,  SU[3], color, g3, False};
\end{verbatim}
The different parts define the names of the superfield, of the gauge group and of the gauge coupling. In addition, the dimension of the gauge group is given. The last entry states, if the gauge indices should be implicit or explicit. The name of the  gaugino component of the vector superfield starts with \verb"f" and the vector boson with \verb"V", i.e. based on the above definition, the gluino is called \verb"fG" and the gluon \verb"VG". The ghost field is named using \verb"g", i.e. \verb"gG".  \\
Chiral superfields are defined by using the list {\tt Fields}, e.g. 
\begin{verbatim}
Fields[[1]] = {{uL, dL},  3, q, 1/6, 2,  3};  
...
Fields[[5]] = {conj[dR],  3, d, 1/3, 1, -3};
\end{verbatim}
The first entry defines the names used for the component fields, then the number of generation and the name for the superfield follows. The automatically created name of the fermionic component starts with \verb"F" and the scalar one with \verb"S": the squarks are named \verb"SuL", \verb"SdL" or \verb"SdR", while the quarks are \verb"FuL", \verb"FdL" and \verb"FdR".\\
After the superfield name, the representation with respect to the gauge groups defined by {\tt Gauge} is assigned.  The transformation of an irreducible representation \(r\) under a given gauge group is in most cases fixed by its dimension \(D\). Therefore, it is sufficient to assign only \(D\) if it is unique.  Otherwise, the Dynkin labels of \(r\) have to be given as  additional input. \\
For all gauge groups, the generators of all appearing representations are needed in order to write the kinetic part of the Lagrangian and the D-terms. In this context, all generators for non-fundamental representations are written as tensor products in \SARAH. Furthermore, the eigenvalues \(C_2(r)\) of the quadratic Casimir as well as the Dynkin index \(I(r)\) are needed. The performed steps to obtain these results are given in \ref{sec:irrep}.
\subsubsection{Superpotential}
The superpotential is defined in a compact form using the variable {\tt SuperPotential}: 
\begin{verbatim}
SuperPotential = {{{Coefficient,Parameter,(Contraction)},
                      {Particle 1, Particle 2, Particle 3} }, ...}
\end{verbatim}
Each term of the superpotential is defined by two lists. The second list assigns all involved fields by using the superfield names. The first list is two- or three-dimensional. It defines a numerical coefficient and the name of the coupling. The gauge and generation indices of the involved superfields are automatically contracted by \SARAH. The used contraction can be displayed via
\begin{verbatim}
ShowSuperpotentialContractions; 
\end{verbatim}
Sometimes, there are more possibilities to contract all indices. Thus, it is possible to fix the contraction of each term using the third entry of the first list.  \\

\subsubsection{Symmetry breaking and rotations}
Rotations for all matter and gauge fields as well as the decomposition of scalar fields into their scalar component, pseudo scalar components and VEVs can be performed. All appearing coefficients as well as the names of the rotation matrices to parametrize this change of the basis can be chosen individually. Besides, it is possible to decompose a field carrying a generation index into its different flavors in order to treat them separately. Afterwards, the complete Lagrangian for the new set of eigenstates is calculated. \\
We give here again only the generic syntax for the different tasks and refer to \ref{sec:MSSM_modelfile} for a discussion of the MSSM. The definition of rotations in the matter sector has either the form
\begin{verbatim}
{Old Basis, {New Basis, Mixing Matrix}}
\end{verbatim} 
or
\begin{verbatim}
{{1.Old Basis,2.Old Basis},
         {{1.New Basis,1.Matrix},{2.New basis,2.Matrix}}}
\end{verbatim} 
depending on whether the mass matrix is hermitian or not. The decomposition of scalar fields is done via
\begin{verbatim}
{Scalar, {{VEV, 1.Coeff.}, {Pseudoscalar, 2.Coeff.},{Scalar,3.Coeff.}};
\end{verbatim}
 Finally, the syntax for the flavor decomposition of fields is
\begin{verbatim}
{Field, {Name for 1.Generation, Name for 2.Generation, ...}};
\end{verbatim}
In principle, these steps can be repeated as often as needed. Therefore, it is no problem to go first to the SCKM basis and afterwards to the mass eigenbasis. GUT theories involving several symmetry breakings can be treated in the same way. The information of all intermediate steps is saved. Hence, it is possible to calculate the vertices or masses of all eigenstates without the necessity of a new model file or a new evaluation of the model. 
\subsubsection{Effective or non-supersymmetric theories}
It is easy in \SARAH to integrate particles out of the spectrum to get an effective theory, or just to delete some particles to get a non-supersymmetric limit of the model. Integrating out particles happens by adding them to the list {\tt IntegrateOut} 
\begin{verbatim}
IntegrateOut = {Particle 1, Particle 2, ...}
\end{verbatim} 
If particles are integrated out, all higher dimensional operators up to dimension 6 are calculated. Deleting particles can be done in the same way as integrating them out. The corresponding list is called {\tt DeleteParticle}.  The difference is that there are no effective operators calculated. Deleting is therefore, of course, faster and should be used if the higher dimensional operators are not needed. 
\subsubsection{Example: Model file for the MSSM}
\label{sec:MSSM_modelfile}
We show in the following the implementation of the MSSM in \SARAH. Our conventions are discussed in \ref{sec:MSSM}.
\begin{enumerate}
\item The gauge sector is \(U(1)\times SU(2)\times SU(3)\) and is just defined by adding the corresponding vector superfields. 
\begin{verbatim}
Gauge[[1]]={B,   U[1], hypercharge, g1, False};
Gauge[[2]]={WB, SU[2], left,        g2, True};
Gauge[[3]]={G,  SU[3], color,       g3, False};
\end{verbatim}
\item The doublet superfields \(\hat{q},\hat{l}, \hat{H}_d, \hat{H}_u\) are added by 
\begin{verbatim}
Fields[[1]] = {{uL,  dL},  3, q,   1/6, 2, 3};  
Fields[[2]] = {{vL,  eL},  3, l,  -1/2, 2, 1};
Fields[[3]] = {{Hd0, Hdm}, 1, Hd, -1/2, 2, 1};
Fields[[4]] = {{Hup, Hu0}, 1, Hu,  1/2, 2, 1};
\end{verbatim}
\item The singlet superfields \(\hat{d}, \hat{u}, \hat{e}\) are added by
\begin{verbatim}
Fields[[5]] = {conj[dR], 3, d,  1/3, 1, -3};
Fields[[6]] = {conj[uR], 3, u, -2/3, 1, -3};
Fields[[7]] = {conj[eR], 3, e,    1, 1,  1};
\end{verbatim}
\item The  superpotential of the MSSM is
\begin{equation}
\label{superpotential_MSSM}
W =  \hat{q} Y_u \hat{u} \hat{H}_u -  \hat{q} Y_d \hat{d} \hat{H}_d  - \hat{l} Y_e \hat{e} \hat{H}_d  +\mu \hat{H}_u \hat{H}_d
\end{equation}
and given in \SARAH by
\begin{verbatim}
SuperPotential = { {{1, Yu},{u,q,Hu}}, {{-1,Yd},{d,q,Hd}},
                   {{-1,Ye},{e,l,Hd}}, {{1,\[Mu]},{Hu,Hd}}  };
\end{verbatim}
\item There are two different sets of eigenstates: the gauge eigenstates before EWSB and the mass eigenstates after EWSB. The internal names are
\begin{verbatim}
NameOfStates={GaugeES, EWSB};
\end{verbatim}
\item The gauge fixing terms for the unbroken gauge groups are
\begin{verbatim}
DEFINITION[GaugeES][GaugeFixing]=
  { {Der[VWB],  -1/(2 RXi[W])},
    {Der[VG],   -1/(2 RXi[G]) }};	
\end{verbatim}
This has the same meaning as
\begin{equation}
\La_{GF} = -\frac{1}{2 \xi_W}|\partial_\mu W^{\mu,i}|^2  -\frac{1}{2 \xi_g}|\partial_\mu g^{\mu,i}|^2
\end{equation}
\item The vector bosons and gauginos rotate after EWSB as follows
\begin{verbatim}
DEFINITION[EWSB][GaugeSector]= 
{ {VWB, {1,{VWm, 1/Sqrt[2]}, {conj[VWm],1/Sqrt[2]}},
        {2,{VWm,-I/Sqrt[2]}, {conj[VWm],I/Sqrt[2]}},
        {3,{VP,Sin[ThetaW]}, {VZ,    Cos[ThetaW]}}},
  {VB,  {1,{VP,Cos[ThetaW]}, {VZ,   -Sin[ThetaW]}}},
  {fWB, {1,{fWm, 1/Sqrt[2]}, {fWp       1/Sqrt[2]}}, 
        {2,{fWm,-I/Sqrt[2]}, {fWp,      I/Sqrt[2]}},
        {3,{fW0,        1}}}                                                         
      };   
\end{verbatim}
This is the common mixing of vector bosons and gauginos after EWSB, see \ref{sec:RotGaugeMSSM}.
\item The neutral components of the scalar Higgs receive VEVs \(v_d\)/\(v_u\) and split in scalar and pseudo scalar components according to \ref{sec:VEVsMSSM}. This is added to \SARAH by
\begin{verbatim}
DEFINITION[EWSB][VEVs]= 
{{SHd0, {vd, 1/Sqrt[2]}, {sigmad, I/Sqrt[2]},{phid, 1/Sqrt[2]}},
 {SHu0, {vu, 1/Sqrt[2]}, {sigmau, I/Sqrt[2]},{phiu, 1/Sqrt[2]}}};
\end{verbatim}
\item The particles mix after EWSB to new mass eigenstates
\begin{verbatim}
DEFINITION[EWSB][MatterSector]= 
{{{SdL, SdR           }, {Sd, ZD}},
 {{SuL, SuR           }, {Su, ZU}},
 {{SeL, SeR           }, {Se, ZE}},
 {{SvL                }, {Sv, ZV}},
 {{phid, phiu         }, {hh, ZH}},
 {{sigmad, sigmau     }, {Ah, ZA}},
 {{SHdm, conj[SHup]   }, {Hpm,ZP}},
 {{fB, fW0, FHd0, FHu0}, {L0, ZN}}, 
 {{{fWm, FHdm}, {fWp, FHup}}, {{Lm,Um},  {Lp,Up}}},
 {{{FeL},       {conj[FeR]}}, {{FEL,ZEL},{FER,ZER}}},
 {{{FdL},       {conj[FdR]}}, {{FDL,ZDL},{FDR,ZDR}}},
 {{{FuL},       {conj[FuR]}}, {{FUL,ZUL},{FUR,ZUR}}} }; 
\end{verbatim}
This defines the mixings to the mass eigenstates described in \ref{sec:RotMatterMSSM}. 
\item The new gauge fixing terms after EWSB are
\begin{verbatim}
DEFINITION[EWSB][GaugeFixing]= 
{{Der[VP],                               - 1/(2 RXi[P])},	
 {Der[VWm]+ I Mass[VWm] RXi[W] Hpm[{1}], - 1/(RXi[W])},
 {Der[VZ] - Mass[VZ] RXi[Z] Ah[{1}],     - 1/(2 RXi[Z])},
 {Der[VG],                               - 1/(2 RXi[G])}};
\end{verbatim}
Based on this definition, \(A^0_1\) and \(H^\pm_1\) are recognized in all calculations as Goldstone bosons.
\item No particles should be integrated out or deleted
\begin{verbatim}
IntegrateOut={};
DeleteParticles={};
\end{verbatim}
\item The Dirac spinors for the mass eigenstates are 
\begin{verbatim}
dirac[[1]] = {Fd,  FdL, FdR};
dirac[[2]] = {Fe,  FeL, FeR};
dirac[[3]] = {Fu,  FuL, FuR};
dirac[[4]] = {Fv,  FvL, 0};
dirac[[5]] = {Chi, L0, conj[L0]};
dirac[[6]] = {Cha, Lm, conj[Lp]};
dirac[[7]] = {Glu, fG, conj[fG]};
\end{verbatim}
\end{enumerate}

\subsection{Parameter and particle files}
\subsubsection{Parameter file}
Additional properties and information about the parameters and particles of a model are saved in the files \verb"parameters.m" and \verb"particles.m". An entry in the parameter file looks like
\begin{verbatim}
{Yu, { LaTeX -> "Y^u",
       Real -> True,
       Form -> Diagonal,
       Dependence ->  None, 
       Value -> None, 
       LesHouches -> Yu
             }}
\end{verbatim}
and contains information about the numerical value ({\tt Value} \(\rightarrow\) number), the position in a LesHouches accord file ({\tt LesHouches} \(\rightarrow\) position) or the dependence on other parameters ({\tt Dependence} \(\rightarrow\) equation). Also simplifying assumptions can be made: it can be defined that parameters contain only real entries ({\tt Real \(\rightarrow\) True}) or that the parameter is diagonal ({\tt Form \(\rightarrow\) Diagonal}). Also a \LaTeX{} name can be given ({\tt LaTeX} \(\rightarrow\) name). Furthermore,   the GUT normalization can be assigned ({\tt GUTnormalization} \(\rightarrow\) value) for the gauge couplings of an  \(U(1)\) gauge groups. 
\subsubsection{Particle file}
\label{sec:ParametersFile}
The particles file contains entries like 
\begin{verbatim}
{Su ,  {  RParity -> -1,
          PDG ->  {1000002,2000002,1000004,2000004,1000006,2000006},
          Width -> Automatic,
          Mass -> Automatic,
          FeynArtsNr -> 13,   
          LaTeX -> "\\tilde{u}",
          OutputName -> "um" }},   
\end{verbatim}
and defines properties of all particles such as the $R$-parity ({\tt RParity} \(\rightarrow\) number) or the mass ({\tt Mass} \(\rightarrow\) value or {\tt Automatic}). {\tt Automatic} means that for the output for \FeynArts or \CalcHep not a fixed numerical value is used, but that the masses are calculated using tree-level relations. In addition, the PDG code is given ({\tt PDG} \(\rightarrow\) number), the number for the particle class used in the \FeynArts model file can be fixed ({\tt FeynArts} \(\rightarrow\) number) and the name in \LaTeX{} form is given ({\tt LaTeX} \(\rightarrow\) name). If a \CalcHep or \CompHep model file should be written, it is also helpful to define an appropriate name in this context ({\tt OutputName} \(\rightarrow\) name). 
\subsubsection{Global definitions}
It is also possible to define global properties for parameters or particles which are present in more than one model file. These properties are afterwards used for all models. The global definitions are saved in the files \verb"particles.m" and \verb"parametes.m" directly in the main model directory. For each parameter or particle, an entry like 
\begin{verbatim}
{{        Descriptions -> "Up Squark", 
          RParity -> -1,
          PDG ->  {1000002,2000002,1000004,2000004,1000006,2000006},
          Width -> Automatic,
          Mass -> Automatic,
          FeynArtsNr -> 13,   
          LaTeX -> "\\tilde{u}",
          OutputName -> "um" }},   
\end{verbatim}
can be added. In particular, the entry \verb"Description" is important. This should be an unique identifier for each particle or parameter. This identifier can later on be used in the different files of the different models, e.g.     
\begin{verbatim}
{Su ,  {  Descriptions -> "Up Squark"}},   
\end{verbatim}
Of course, it is also possible to overwrite some of the global definitions by defining them locally, too. For instance, to use \verb"u" instead of \verb"um" as output name in a specific model, the entry should be changed to  
\begin{verbatim}
{Su ,  {  Descriptions -> "Up Squark",
          OutputName -> "u" }},   
\end{verbatim}
in the corresponding particle file of the model.
\section{Verification}
\paragraph*{Tree-level results}  We have checked the model files generated with \SARAH  for the MSSM against the existing files of \FeynArts and \CalcHep. The checks happened as well at vertex level as for complete processes. We have compared the numerical value of each vertex for different sets of parameters and all possible combinations of generations (more than 5000). In addition we have calculated several \(1\rightarrow 2\) and \(2\rightarrow 2\) processes with the old and new model files. Similar checks have been done for the vertices of the NMSSM against the model file of \CalcHep. More information about the verification of the tree-level results is given \cite{sarah1}.
\paragraph*{One-loop self-energies and tadpole equations} We have compared the analytical expressions of the self-energies calculated by \SARAH for the MSSM with the results of \cite{Pierce:1996zz}. In addition, we have compared the numerical values for the one-loop corrected masses with the results of {\tt SPheno} \cite{SPheno}. Furthermore, the results for the NMSSM were numerically checked against the routines provided by the authors of \cite{Degrassi:2009yq}. Both results were in complete agreement. A detailed discussion about the results for the NMSSM obtained by \SARAH is also given in \cite{Staub:2010ty}.
\paragraph*{Verification of the calculated RGEs}
We have compared the analytical results for the one- and two-loop RGEs calculated by \SARAH for the MSSM with \cite{Martin:1993zk} and for the NMSSM with \cite{Ellwanger}. The only difference has been in the NMSSM the two-loop RGE of \(A_\lambda\). A second calculation by authors of  \cite{Ellwanger} have confirmed the first result of \SARAH. \\
In addition, we have checked a model containing non-fundamental representations: the \(SU(5)\) inspired Seesaw II model of \cite{Borzumati} and \cite{Rossi}. It is known that there are discrepancies of the RGEs given in these two papers. Our result fully agrees with  \cite{Borzumati}. \\
Also numerically checks have been done by comparing the RGEs for the MSSM with the RGEs implemented in {\tt SPheno} \cite{SPheno}. Both sets of RGEs are in full agreement.

\section*{Acknowledgments}
This work is supported by the DFG Graduiertenkolleg GRK-1147 and by the DAAD, project number D/07/13468. We thank Bj\"orn Hermann and Avelino Vincente for their checks of the output and useful suggestions, Ritesh Singh for proof-reading the manuscript and Werner Porod for fruitful discussions. In addition, we thank Nicole Schatz for her support during all the work.    

\begin{appendix}

\section{Gauge anomalies}
\label{sec:GaugeAnomaly}
Before \SARAH starts the calculation of the Lagrangian it checks the model for the different triangle anomalies. These anomalies can involve diagrams with three external gauge bosons belonging to the same \(U(1)\) or \(SU(N)\) gauge group. To be anomaly free, the sum over all internal fermions has to vanish
\begin{eqnarray}
 U(1)^3_i &:&  \sum_n {Y^i_n}^3 = 0 \thickspace , \\
 SU(N)^3_i &:&  \sum_n \mbox{Tr}(T^i_n T^i_n T^i_n) = 0 \thickspace .
 \end{eqnarray}
We label the different gauge groups with the indices \(i,j,k\). \(Y^i_n\) is the charge of particle \(n\) under the abelian gauge group \(i\) while \(T^i_n\) is the generator with respect to a non-abelian gauge group.\\
Combinations of two different gauge groups are possible, if one group is an \(U(1)\). Hence, another condition for the absence of anomalies is
\begin{equation}
 U(1)_i\times SU(N)^2_j  :  \sum_n Y^i_n\, \mbox{Tr}(T^j_n T^j_n) = 0  \thickspace .
\end{equation}
If more than one \(U(1)\) gauge group are present, anomalies can be generated by  two or three different \(U(1)\) gauge bosons as external fields, too. Therefore, it has to be checked, that 
\begin{eqnarray}
 U(1)_i\times U(1)_j^2 &:& \thinspace  \sum_n  Y^i_n {Y^j_n}^2 = 0 \thickspace , \\
 U(1)_i\times U(1)_j\times U(1)_k &:& \thinspace  \sum_n  Y^i_n Y^j_n  Y^k_n= 0
\end{eqnarray}
holds. In addition, it has to be checked that there is an even number of \(SU(2)\) doublets. This is necessary for a model in order to be free of the Witten anomaly \cite{Witten:1982fp}. If one condition is not fulfilled, a warning is given by \SARAH but the model can be evaluated anyway.
\section{Calculation of the Lagrangian of supersymmetric models}
\label{sec:Lagrangian}
We describe in this section the calculation of the complete Lagrangian for a supersymmetric model based on the superpotential and the gauge structure. 
\paragraph*{Interactions of chiral superfields}
If we call the superpotential for a given theory \(W\) and use \(\phi_i\) for the scalar and \(\psi_i\) for the fermionic component of a chiral supermultiplet, the matter interactions can by derived by
\begin{equation}
\label{eq:MatterInteractions}
\La_Y = - \frac{1}{2} W^{ij} \psi_i \psi_j + \mbox{h.c.} \thickspace, \hspace{1cm} \La_F = F^{* i} F_i + \mbox{h.c.}
\end{equation}
with 
\begin{equation}
\label{eq:FTerms}
W^{ij} = \frac{\delta^2}{\delta \phi_i \delta \phi_j} W \hspace{1cm} \mbox{and} \hspace{1cm} F^i = - W^{* i} = \frac{\delta W}{\delta \phi_i} \thickspace .
\end{equation}
The first term of eq. (\ref{eq:MatterInteractions}) describes  the interaction of two fermions with one scalar, while the second term forms the so called F-terms which describe  four-scalar interactions.  

\paragraph*{Interactions of vector superfields}
We name  the spin-\(\frac{1}{2}\) component of a vector supermultiplet \(\lambda\) and  the spin-1 component \(A^\mu\). The most general Lagrangian only involving these fields is
\begin{equation}
\label{eq:LagVS}
\La = - \frac{1}{4} F^a_{\mu\nu} F^{\mu\nu a} - i \lambda^{\dagger a } \bar{\sigma}^\mu D_\mu \lambda^a
\end{equation}
with the field strength
\begin{equation}
\label{eq:FieldStrength}
F_{\mu\nu}^a = \partial_\mu A_\nu^a - \partial_\nu A_\mu^a + g f^{abc} A_\mu^b A_\nu^c \thickspace ,
\end{equation}
and the covariant derivative
\begin{equation}
D_\mu \lambda^a = \partial_\mu \lambda^a + g f^{abc} A_\mu^b \lambda^c \thickspace .
\end{equation}
Here, \(f^{abc}\) is the structure constant of the gauge group. Plugging eq.~(\ref{eq:FieldStrength}) in the first term of eq.~(\ref{eq:LagVS}) leads to self-interactions of three and four gauge bosons
\begin{equation}
\La_{V} = - \frac{1}{4} (\partial_\mu A_\nu^a - \partial_\nu A_\mu^a)  g f^{abc} A^{\mu,b} A^{\nu,c}  - \frac{1}{4} g^2 (f_{abc} A_\mu^b A_\nu^c) (f^{ade} A^{\mu,e} A^{\nu,e})  \thickspace .
\end{equation}
The second term of eq.~(\ref{eq:LagVS})  describes the  interactions between vector bosons and gauginos.
\paragraph*{Supersymmetric gauge interactions}
The parts of the Lagrangian with both chiral and vector superfields are the kinetic terms for the fermions and scalars
\begin{equation}
\La_{kin} = - D^\mu \phi^{*i} D_\mu \phi_i  - i \psi^{\dagger i } \bar{\sigma}^\mu D_\mu \psi_i 
\end{equation}
as well as the interaction between a gaugino and a matter fermion and scalar
\begin{equation}
\La_{GFS} = - \sqrt{2} g (\phi^* T^a \psi) \lambda^a + \mbox{h.c.} \thickspace .
\end{equation} 
Here, \(T^a\) are the fundamental generators of the gauge group. Furthermore, the covariant derivatives are 
\begin{eqnarray}
D_\mu \phi_i &=& \partial_\mu \phi_i - ig A^a_\mu (T^a \phi)_i  \thickspace ,\\
D_\mu \phi^{*i} &=& \partial_\mu \phi^{*i} + ig A^a_\mu (\phi^* T^a)^i \thickspace , \\
D_\mu \psi_i &=& \partial_\mu \psi_i - i g A^a_\mu (T^a \psi)_i \thickspace,
\end{eqnarray}
In addition, the D-terms are defined by
\begin{equation}
\La_D =  \frac{1}{2} D^a D^a \thickspace .
\end{equation}
The solution of the equations of motion for the auxiliary fields leads to 
\begin{equation}
\label{eq:DTerms}
D^a = - g (\phi^* T^a \phi) \thickspace .
\end{equation}
\paragraph*{Soft-breaking terms} SUSY must be a broken. This can be parametrized by adding soft-breaking terms to the Lagrangian. The possible terms are the mass terms for all scalar matter fields and gauginos
\begin{equation}
\La_{SB} = - m_{\phi_i}^2 \phi_i \phi_i^* - \frac{1}{2} M_{\lambda_i} \lambda_i \lambda_i
\end{equation}
as well as soft-breaking interaction corresponding to the superpotential terms
\begin{equation}
\La_{Soft,W} = T \phi_i \phi_j \phi_k + B \phi_i \phi_j + S \phi_i \thickspace .
\end{equation}
\paragraph*{Gauge fixing terms and ghost interactions}
\label{GaugeFixing}
The Lagrangian of a theory without further restrictions is invariant under a general gauge transformation. This invariance leads to severe problems in the quantization of the theory as can be seen in the divergence of functional integrals. Therefore, it is in necessary to add gauge fixing terms to break this gauge invariance. \\
The general form of the gauge fixing Lagrangian is
\begin{equation}
\La_{GF} = - \frac{1}{2} \sum_a|f(x)^a|^2 \thickspace .
\end{equation} 
\(f_a\) can be a function of partial derivatives of a gauge boson and a Goldstone boson. The corresponding ghost terms of the ghost fields \(\bar{\eta}\) and \(\eta\)  are
\begin{equation}
\La_{Ghost} = - \bar{\eta}_a (\delta f^a) \thickspace. 
\end{equation}
Here, \(\delta\) assigns the operator for a BRST transformation. For an unbroken gauge symmetry, the gauge fixing terms in the often chosen \(R_\xi\)-gauge  are
\begin{equation}
\La_{GF} = - \frac{1}{2 R_\xi} \sum_a \left(\partial^\mu V_\mu^a \right)^2  \thickspace.
\end{equation}
Here, \(V_\mu\) are the gauge boson of the unbroken gauge group. It is often common to choose a distinct value for \(R_\xi\). The most popular gauges are the unitary gauge \(R_\xi \rightarrow \infty\) and the Feynman-'t Hooft-gauge \(R_\xi = 1\). For broken symmetries, the gauge fixings terms are chosen in a way that the mixing terms between vector bosons and scalars disappear from the Lagrangian. Therefore, the common choice for the gauge fixing Lagrangian for theories with the standard model gauge sector after EWSB is  
\begin{equation}
\label{GFewsb}
\La_{GF, R_\xi} =  - \frac{1}{2 \xi_\gamma} \left( \partial^\mu \gamma_\mu\right)^2 - \frac{1}{2 \xi_Z} \left( \partial^\mu Z_\mu + \xi_Z M_Z G^0 \right)^2 + - \frac{1}{\xi_{W^+}} \left( \partial^\mu W^+_\mu + \xi_{W^+} M_W G^+\right)^2 \thickspace . 
\end{equation}
Here, \(G^0\) and \(G^+\) are the Goldstone bosons, which build the longitudinal component of the massive vector bosons. \\
\section{Calculation of Group Factors}
\label{sec:irrep}
SARAH supports not only chiral superfields in the fundamental representation but in any irreducible representation of \(SU(N)\). In most cases, it is possible to fix the transformation properties of the chiral superfield by stating the corresponding dimension \(D\). If the dimension is not unique, also the Dynkin labels are needed. For calculating kinetic terms and D-terms, it is necessary to derive from the representation the corresponding generators. Furthermore,  the eigenvalues \(C_2\) of the quadratic Casimir for any irreducible representation \(r\) 
\begin{equation}
T^a T^a \phi(r) = C_2(r) \phi(r)
\end{equation}
as well as the Dynkin index \(I\)
\begin{equation}
Tr(T^a T^b) \phi(r) = I \delta_{a b} \phi(r)
\end{equation}
are needed for the calculation of the RGEs. All of that is derived by \SARAH due to the technique of Young tableaux. The following steps are evolved:
\begin{enumerate}
\item The corresponding Young tableau fitting to the dimension \(D\) is calculated using the hook formula:
\begin{equation}
D = \Pi_i \frac{N + d_i}{h_i} \thickspace,
\end{equation}
\(d_i\) is the distance of the \(i.\) box to the left upper corner and \(h_i\) is the hook of that box. \\
\item The vector for the highest weight \(\Lambda\) in Dynkin basis is extracted from the tableau.
\item The fundamental weights for the given gauge group are calculated.
\item The value of \(C_2(r)\) is calculated using the Weyl formula
\begin{equation}
C_2(r) = ( \Lambda, \Lambda + \rho) \thickspace ,
\end{equation}
\(\rho\) is the Weyl vector. 
\item The Dynkin index \(I(r)\) is calculated from \(C_2(r)\). For this step, the value for the fundamental representation is normalized to \(\frac{1}{2}\). 
\begin{equation}
I(r) = C_2(r) \frac{D(r)}{D(G)} \thickspace ,
\end{equation}
with \(D(G)\) as dimension of the adjoint representation.
\item The number of co- and contra-variant indices is extracted from the Young tableau. With this information, the generators are written as tensor product.
\end{enumerate}
The user can calculate this information independently from the model using the new command
\begin{verbatim}
CheckIrrepSUN[Dim,N]
\end{verbatim}
\verb"Dim" is the dimension of the irreducible representation and \verb"N" is the dimension of the \(SU(N)\) gauge group. 
The result is a vector containing the following information: (i) repeating the dimension of the field, (ii) number of covariant indices, (iii) number of contravariant indices, (iv) value of the quadratic Casimir \(C_2(r)\), (v) value of the Dynkin index \(I(r)\), (vi) Dynkin labels for the highest weight.
\paragraph*{Examples}
\begin{enumerate}
\item {\bf Fundamental representation}. The properties of a particle, transforming under the fundamental representation of \(SU(3)\) are calculated via {\tt CheckIrrepSUN[3,3]}. The output is the well known result
\begin{verbatim}
{3, 1, 0, 4/3, 1/2, {1, 0}}
\end{verbatim}
\item {\bf Adjoint representation}. The properties of a field transforming as {\bf 24} of \(SU(5)\) are calculated by
{\tt  CheckIrrepSUN[24,5] }. The output will be
\begin{verbatim}
{24, 1, 1, 5, 5, {1, 0, 0, 1}}
\end{verbatim}
\item {\bf  Different representations with the same dimension}.
The {\bf{70}} under \(SU(5)\) is not unique. Therefore, {\tt CheckIrrepSUN[\{70, \{0, 0, 0, 4\}\}, 5] }  returns
\begin{verbatim}
{70, 0, 4, 72/5, 42, {0, 0, 0, 4}} 
\end{verbatim}
while {\tt CheckIrrepSUN[\{70, \{2, 0, 0, 1\}\}, 5] } leads to
\begin{verbatim}
{70, 2, 1, 42/5, 49/2, {2, 0, 0, 1}} 
\end{verbatim}
\end{enumerate}
\section{Conventions and generic expressions}
\label{app:formulas}
\subsection{Renormalization group equations}
\label{app:RGE_con}
We summarize in this section the used equations for the calculation of the one- and two-loop RGEs in \SARAH. These equations are extensively discussed in literature, see e.g. \cite{Martin:1993zk,Jack:1998iy, Jack:1994rk, Jones:1984cx, West:1984dg, Jack:1997eh, Yamada:1993uh, Yamada:1993ga, Yamada:1994id}.  \\
For a  general $N=1$ supersymmetric gauge theory with superpotential  
\begin{equation}
 W (\phi) = L_i \phi_i + \frac{1}{2}{\mu}^{ij}\phi_i\phi_j + \frac{1}{6}Y^{ijk}
\phi_i\phi_j\phi_k \thickspace ,
\end{equation}
the  soft SUSY-breaking scalar terms are given by
\begin{equation}
V_{\hbox{soft}} = \left(S^i \phi_i +   \frac{1}{2}b^{ij}\phi_i\phi_j
+ \frac{1}{6}h^{ijk}\phi_i\phi_j\phi_k +\hbox{c.c.}\right)
+(m^2)^i{}_j\phi_i\phi_j^* + \frac{1}{2} M_\lambda \lambda_a \lambda_a \thickspace.
\end{equation}
The anomalous dimensions are given by 
\begin{align}
 \gamma_i^{(1)j} = & \frac{1}{2} Y_{ipq} Y^{jpq} - 2 \delta_i^j g^2 C(i) \thickspace ,\\
 \gamma_i^{(2)j}  = &  -\frac{1}{2} Y_{imn} Y^{npq} Y_{pqr} Y^{mrj} + g^2 Y_{ipq} Y^{jpq} [2C(p)- C(i)] \nonumber \\
 & \; \;  + 2 \delta_i^j g^4 [ C(i) S(R)+ 2 C(i)^2 - 3 C(G) C(i)] \thickspace ,
\end{align}
and the \(\beta\)-functions for the gauge couplings are given by
\begin{align}
 \beta_g^{(1)}  =  &  g^3 \left[S(R) - 3 C(G) \right] \thickspace , \\
 \beta_g^{(2)}  =  &  g^5 \left\{ - 6[C(G)]^2 + 2 C(G) S(R) + 4 S(R) C(R) \right\}
    - g^3 Y^{ijk} Y_{ijk}C(k)/D(G) \thickspace .
\end{align}
Here, \(C(i)\) is the quadratic Casimir for a specific superfield and $C(R),C(G)$ are the quadratic Casimirs for the matter and adjoint  representations, respectively. \(D(G)\) is the dimension of the adjoint representation.  \\
The corresponding RGEs are defined as
\begin{equation}
\frac{d}{dt} g  =   \frac{1}{16\pi^2} \beta_g^{(1)} +  \frac{1}{(16\pi^2)^2} \beta_g^{(2)} \thickspace.
\end{equation}
Here, we used \(t=\ln Q\), where \(Q\) is the renormalization scale. The $\beta$-functions for the superpotential parameters can be obtained by using superfield technique. The obtained expressions are 
\begin{eqnarray}
 \beta_Y^{ijk} &= & Y^{ijp} \left [
 \frac{1}{16\pi^2}\gamma_p^{(1)k} +
 \frac{1}{(16\pi^2)^2}  \gamma_p^{(2)k} \right ]
+ (k \leftrightarrow i) + (k\leftrightarrow j) \thickspace , \\
 \beta_{\mu}^{ij} &= & \mu^{ip} \left [
 \frac{1}{16\pi^2}\gamma_p^{(1)j} +
 \frac{1}{(16\pi^2)^2} \gamma_p^{(2)j} \right ]
+ (j \leftrightarrow i) \thickspace , \\
\beta_L^i & = & L^{p} \left [
 \frac{1}{16\pi^2}\gamma_p^{(1)i} +
 \frac{1}{(16\pi^2)^2}  \gamma_p^{(2)i} \right ] \thickspace .
\end{eqnarray}
The expressions for trilinear, soft-breaking terms are
\begin{align}
\frac{d}{dt} h^{ijk}  =  & \frac{1}{16\pi^2} \left [\beta^{(1)}_h\right ]^{ijk} +  \frac{1}{(16\pi^2)^2} \left [\beta^{(2)}_h\right ]^{ijk} \thickspace ,
\end{align}
with
\begin{align}
\left [\beta^{(1)}_h\right ]^{ijk}   =
 & \frac{1}{2} h^{ijl} Y_{lmn} Y^{mnk}
+ Y^{ijl} Y_{lmn} h^{mnk} - 2 \left (h^{ijk} - 2 M Y^{ijk}  \right )
g^2  C(k) \nonumber \\
& + (k \leftrightarrow i) + (k \leftrightarrow j) \thickspace , \\
\left [\beta^{(2)}_h\right ]^{ijk}   =
 &
-\frac{1}{2} h^{ijl} Y_{lmn} Y^{npq} Y_{pqr} Y^{mrk} \nonumber \\
& - Y^{ijl} Y_{lmn} Y^{npq} Y_{pqr} h^{mrk}
- Y^{ijl} Y_{lmn} h^{npq} Y_{pqr} Y^{mrk} \nonumber \\
& + \left ( h^{ijl} Y_{lpq} Y^{pqk} +  2 Y^{ijl} Y_{lpq} h^{pqk}
- 2 M Y^{ijl} Y_{lpq} Y^{pqk} \right ) g^2\left[ 2 C(p) - C(k) \right ]  \nonumber \\
& + \left (2h^{ijk} - 8 M Y^{ijk} \right )
g^4 \left [  C(k)S(R)+ 2 C(k)^2  - 3 C(G)C(k)\right ]\nonumber  \\
&+ (k \leftrightarrow i) + (k \leftrightarrow j)   \thickspace .
\end{align}
For the bilinear soft-breaking parameters, the expressions read
\begin{align}
\frac{d}{dt} b^{ij}  =  &
 \frac{1}{16\pi^2}\left [\beta^{(1)}_b \right ]^{ij} +
 \frac{1}{(16\pi^2)^2} \left [\beta^{(2)}_b \right ]^{ij} \thickspace , 
\end{align}
with
\begin{align}
\left [\beta^{(1)}_b \right ]^{ij}   = 
&\frac{1}{2} b^{il} Y_{lmn} Y^{mnj} +\frac{1}{2}Y^{ijl} Y_{lmn} b^{mn}
+ \mu^{il} Y_{lmn} h^{mnj}
- 2 \left (b^{ij} - 2 M \mu^{ij} \right )g^2 C(i)  \nonumber \\ 
& + (i \leftrightarrow j) \thickspace , \\
\left [\beta^{(2)}_b \right ]^{ij}   = 
& -\frac{1}{2} b^{il} Y_{lmn} Y^{pqn} Y_{pqr} Y^{mrj}
-\frac{1}{2} Y^{ijl} Y_{lmn} b^{mr} Y_{pqr} Y^{pqn} \nonumber \\
&-\frac{1}{2} Y^{ijl} Y_{lmn} \mu^{mr} Y_{pqr} h^{pqn}
- \mu^{il} Y_{lmn} h^{npq} Y_{pqr}  Y^{mrj} \nonumber \\
& - \mu^{il} Y_{lmn} Y^{npq}  Y_{pqr} h^{mrj}
+ 2 Y^{ijl} Y_{lpq} \left ( b^{pq} - \mu^{pq} M \right ) g^2 C(p)
 \nonumber \\
& + \left ( b^{il} Y_{lpq} Y^{pqj} + 2 \mu^{il} Y_{lpq} h^{pqj}
- 2 \mu^{il} Y_{lpq} Y^{pqj} M \right )
g^2 \left[ 2 C(p) - C(i) \right ]  \nonumber \\
& +  \left ( 2 b^{ij} - 8 \mu^{ij} M\right )
g^4 \left [ C(i)S(R)+ 2 C(i)^2- 3 C(G)C(i)   \right ] \nonumber \\
&+ (i \leftrightarrow j)  \thickspace , 
\end{align}
Finally, the RGEs for the linear soft-breaking parameters are
\begin{align}
\frac{d}{dt} S^{i}  =  &
 \frac{1}{16\pi^2}\left [\beta^{(1)}_S \right ]^{i} +
 \frac{1}{(16\pi^2)^2} \left [\beta^{(2)}_S \right ]^{i} \thickspace ,
\end{align}
with
\begin{align}
\left [\beta^{(1)}_S \right]^i  =&
\frac{1}{2}Y^{iln}Y_{pln}S^{p}
+L^{p}Y_{pln}h^{iln}
+\mu^{ik}Y_{kln}B^{ln}+2Y^{ikp}(m^2)_{p}^l\mu_{kl}
+h^{ikl}B_{kl} \thickspace ,\\
\left [\beta^{(2)}_S \right]^i  = &
2g^2C(l)Y^{ikl}Y_{pkl}S^{p} -\frac{1}{2}Y^{ikq}Y_{qst}Y^{lst}Y_{pkl}S^{p}
-4g^2C(l)(Y^{ikl}M-h^{ikl}) Y_{pkl}L^{p}  \nonumber \\
& -\big[Y^{ikq}Y_{qst}h^{lst}Y_{pkl} +h^{ikq}Y_{qst}Y^{lst}Y_{pkl}\big]L^{p}
-4g^2C(l)Y_{jnl} (\mu^{nl}M-B^{nl})\mu^{ij} 
\nonumber \\
& -\big[Y_{jnq}h^{qst}Y_{lst}\mu^{nl} +Y_{jnq}Y^{qst}Y_{lst}B^{nl}\big]\mu^{ij}  
+4g^2C(l)(2Y^{ikl}\mu_{kl}|M|^2-Y^{ikl}B_{kl}M
\nonumber \\
&-h^{ikl}\mu_{kl} M^* +h^{ikl}B_{kl}+Y^{ipl}(m^2)_{p}^k\mu_{kl}
+Y^{ikp}(m^2)_{p}^l\mu_{kl}) 
\nonumber \\
 &-\Big[Y^{ikq}Y_{qst}h^{lst}B_{kl}+h^{ikq}Y_{qst}Y^{lst}B_{kl}
+Y^{ikq}h_{qst}h^{lst}\mu_{kl}  +h^{ikq}h_{qst}Y^{lst}\mu_{kl} \nonumber \\
\nonumber &
+Y^{ipq}(m^2)_{p}^kY_{qst}Y^{lst}\mu_{kl}
+Y^{ikq}Y_{qst}Y^{pst}(m^2)_{p}^l\mu_{kl}
+Y^{ikp}(m^2)_{p}^qY_{qst}Y^{lst}\mu_{kl}
\\ & +2Y^{ikq}Y_{qsp}(m^2)_t^{p}Y^{lst}\mu_{kl} \Big] \thickspace .
\end{align}
With these results, the list of the \(\beta\)-functions for all couplings is complete. Now, we turn to  the RGEs for the gaugino masses, squared masses of scalars and vacuum expectation values. The result for the gaugino masses is 
\begin{align}
 \frac{d}{dt} M = & \frac{1}{16 \pi^2} \beta_M^{(1)}
+ \frac{1}{(16 \pi^2)^2 } \beta_M^{(2)} \thickspace ,
\end{align}
with
\begin{align}
\beta_M^{(1)} = & g^2 \left[ 2 S(R) - 6 C(G) \right] M \thickspace ,
\\
\beta_M^{(2)}
= & g^4\left\{ -24[C(G)]^2 + 8 C(G) S(R) + 16  S(R)C(R)\right\} M\cr
 &\hbox{\hskip 110pt} + 2 g^2 \left [h^{ijk}  - M Y^{ijk}\right] Y_{ijk}
C(k)/D(G) \thickspace .
\end{align}
The one- and two-loop RGEs for the scalar mass parameters read
\begin{align}
\frac{d}{dt} m_{ij}  =  &
 \frac{1}{16\pi^2} \left [\beta^{(1)}_{m^2} \right ]_i^j +
 \frac{1}{(16\pi^2)^2} \left [\beta^{(2)}_{m^2} \right ]_i^j \thickspace , \\
\end{align}
with
\begin{align}
\left [\beta^{(1)}_{m^2} \right ]_i^j  =  &
 \frac{1}{2} Y_{ipq} Y^{pqn} {(m^2)}_n^j
+ \frac{1}{2} Y^{jpq} Y_{pqn} {(m^2)}_i^n + 2 Y_{ipq} Y^{jpr} {(m^2)}_r^q
\nonumber \\
& + h_{ipq} h^{jpq}  - 8 \delta_i^j M M^\dagger g^2 C(i) +
 2 g^2 {\bf t}^{Aj}_i {\rm Tr} [ {\bf t}^A m^2 ] \thickspace , \\
\left [\beta^{(2)}_{m^2} \right ]_i^j  =  &
 -\frac{1}{2} {(m^2)}_i^l Y_{lmn} Y^{mrj} Y_{pqr} Y^{pqn}
 -\frac{1}{2} {(m^2)}^j_l Y^{lmn} Y_{mri} Y^{pqr} Y_{pqn} \nonumber \\
& - Y_{ilm} Y^{jnm} {(m^2)}_r^l Y_{npq} Y^{rpq}
 - Y_{ilm} Y^{jnm} {(m^2)}_n^r Y_{rpq} Y^{lpq} \nonumber  \\
& - Y_{ilm} Y^{jnr} {(m^2)}_n^l Y_{pqr} Y^{pqm}
- 2 Y_{ilm} Y^{jln}  Y_{npq} Y^{mpr} {(m^2)}_r^q \nonumber \\
& - Y_{ilm} Y^{jln} h_{npq} h^{mpq} - h_{ilm} h^{jln} Y_{npq} Y^{mpq}  \nonumber\\
& - h_{ilm} Y^{jln} Y_{npq} h^{mpq} - Y_{ilm} h^{jln} h_{npq} Y^{mpq} \nonumber\\
& + \biggl [{(m^2)}_i^l Y_{lpq} Y^{jpq}
+ Y_{ipq} Y^{lpq} {(m^2)}_l^j + 4 Y_{ipq} Y^{jpl} {(m^2)}_l^q
+  2 h_{ipq} h^{jpq}
\nonumber \\
&  - 2 h_{ipq} Y^{jpq} M -2 Y_{ipq} h^{jpq} M^\dagger
+ 4Y_{ipq} Y^{jpq} M M^\dagger
\biggr ]
g^2 \left [C(p) + C(q)- C(i) \right ] \nonumber \\
& -2 g^2 {\bf t}^{Aj}_i ({\bf t}^A m^2)_r^l Y_{lpq} Y^{rpq}
+ 8 g^4 {\bf t}^{Aj}_i {\rm Tr} [ {\bf t}^A C(r) m^2 ]  \nonumber \\
& + \delta_i^j g^4 M M^\dagger \left [
24C(i) S(R) + 48 C(i)^2 - 72 C(G) C(i) \right ]
\nonumber \\ 
& + 8 \delta_i^j g^4 C(i) ( {\rm Tr} [S(r) m^2] - C(G) M M^\dagger ) \thickspace .
\end{align}
The RGEs for a VEV \(v^i\) are proportional to the anomalous dimension of the chiral superfield whose scalar component receives the VEV
\begin{equation}
 \frac{d}{dt }v^i =  v^{p} \left [
 \frac{1}{16\pi^2}\gamma_p^{(1)i} +
 \frac{1}{(16\pi^2)^2}  \gamma_p^{(2)i} \right ]
\end{equation}

\subsection{One-loop amplitudes for one- and two-point functions}
\label{sec:Integrals}
We used for the calculation of the one-loop self-energies and the one-loop corrections to the tadpoles in \(\DR\)-scheme the scalar functions defined in \cite{Pierce:1996zz}. The basis integrals are
\begin{eqnarray}
A_0(m) &=& 16\pi^2Q^{4-n}\int{\frac{d^nq}{ i\,(2\pi)^n}}{\frac{1}{ q^2-m^2+i\varepsilon}} \thickspace ,\\
B_0(p, m_1, m_2) &=& 16\pi^2Q^{4-n}\int{\frac{d^nq}{ i\,(2\pi)^n}} {\frac{1}{\biggl[q^2-m^2_1+i\varepsilon\biggr]\biggl[
(q-p)^2-m_2^2+i\varepsilon\biggr]}} \thickspace ,
\label{B0 def}
\end{eqnarray}
with the renormalization scale \(Q\). The integrals are regularized by integrating in $n=4-2\epsilon$ dimensions. The result for \(A_0\) is
\begin{equation}
A_0(m)\ =\ m^2\left({\frac{1}{\hat\epsilon}} + 1 - \ln{\frac{m^2}{Q^2}}\right)~,\label{A}
\end{equation}
where $1/\hat\epsilon =1/\epsilon-\gamma_E+\ln 4\pi$. The function $B_0$  has the analytic expression
\begin{equation}
 B_0(p, m_1, m_2) \ =\ {\frac{1}{\hat\epsilon}} - \ln\left(\frac{p^2}{Q^2}\right) - f_B(x_+) - f_B(x_-)~,
\end{equation}
with
\begin{equation}
 x_{\pm}\ =\ \frac{s \pm \sqrt{s^2 - 4p^2(m_1^2-i\varepsilon)}}{2p^2}~,
\qquad f_B(x) \ =\ \ln(1-x) - x\ln(1-x^{-1})-1~,
\end{equation}
and $s=p^2-m_2^2+m_1^2$. All the other, necessary functions can be expressed by $A_0$ and $B_0$. For instance,
\begin{equation}
 B_1(p, m_1,m_2) \ =\ {\frac{1}{2p^2}}\biggl[ A_0(m_2) - A_0(m_1) + (p^2
+m_1^2 -m_2^2) B_0(p, m_1, m_2)\biggr]~,
\end{equation}
and
\begin{eqnarray}
B_{22}(p, m_1,m_2) &=& \frac{1}{6}\ \Bigg\{\,
\frac{1}{2}\biggl(A_0(m_1)+A_0(m_2)\biggr)
+\left(m_1^2+m_2^2-\frac{1}{2}p^2\right)B_0(p,m_1,m_2)\nonumber \\ &&+ \frac{m_2^2-m_1^2}{2p^2}\ \biggl[\,A_0(m_2)-A_0(m_1)-(m_2^2-m_1^2)
B_0(p,m_1,m_2)\,\biggr] \nonumber\\ && +  m_1^2 + m_2^2
-\frac{1}{3}p^2\,\Bigg\}~.
\end{eqnarray}
Furthermore, for the self-energies of vector bosons, it is useful to define
\begin{align}
F_0(p,m_1,m_2) =& A_0(m_1)-2A_0(m_2)- (2p^2+2m^2_1-m^2_2)B_0(p,m_1,m_2)
\ , \\ 
G_0(p,m_1,m_2) =&
(p^2-m_1^2-m_2^2)B_0(p,m_1,m_2)-A_0(m_1)-A_0(m_2)\ ,\\
H_0 (p,m_1,m_2) =& 4B_{22}(p,m_1,m_2) + G(p,m_1,m_2)\ ,\\
\tilde{B}_{22}(p,m_1,m_2) =& B_{22}(p,m_1,m_2) - \frac{1}{4}A_0(m_1) -
\frac{1}{4}A_0(m_2)
\end{align}
%
In addition, several coefficients are involved:
\begin{itemize}
 \item \(c_S\) is the symmetry factor: if the particles in the loop are indistinguishable, the weight of the contribution is only half of the weight in the case of distinguishable particles. If two different charge flows are possible in the loop, the weight of the diagram is doubled.
 \item \(c_C\) is a charge factor: for corrections due to vector bosons in the adjoint representation this is the Casimir of the corresponding group. For corrections due to matter fields this can be, for instance, a color factor for quarks/squarks. For corrections of vector bosons in the adjoint representation this is normally the Dynkin index of the gauge group.
 \item  \(c_R\) is 2 for real fields and Majorana fermions in the loop and 1 otherwise. \\
\end{itemize}
We use in the following \(\Gamma\) for non-chiral interactions and \(\Gamma_L\)/\(\Gamma_R\) for chiral interactions. If two vertices are involved, the interaction of the incoming particle has an upper index 1 and for the outgoing field an upper index 2 is used. 
\subsubsection{One-loop tadpoles}
\begin{enumerate}
\item Fermion loop (generic name in \SARAH: \verb"FFS"):
\begin{equation}
T = 8 c_S c_C m_F \Gamma A_0(m_F^2) 
\end{equation}
\item Scalar loop (generic name in \SARAH: \verb"SSS"):
\begin{equation}
T = - 2 c_S c_C \Gamma A_0(m_S^2) 
\end{equation}
\item Vector boson loop (generic name in \SARAH: \verb"SVV"):
\begin{equation}
T = 6 c_S c_C \Gamma A_0(m_V^2) 
\end{equation}
\end{enumerate}

\subsubsection{One-loop self-energies}
\paragraph{Corrections to fermion}
\begin{enumerate}
\item Fermion-scalar loop (generic name in \SARAH: \verb"FFS"):
\begin{eqnarray*}
\Sigma^S(p^2) &=& m_F c_S c_C c_R \Gamma^1_R \Gamma^{2,*}_L B_0(p^2,m_F^2,m_S^2) \\
\Sigma^R(p^2) &=& - c_S c_C c_R \frac{1}{2} \Gamma^1_R \Gamma^{2,*}_R B_1(p^2,m_F^2,m_S^2) \\
\Sigma^L(p^2) &=& - c_S c_C c_R \frac{1}{2} \Gamma^1_L \Gamma^{2,*}_L B_1(p^2,m_F^2,m_S^2) 
\end{eqnarray*}
\item Fermion-vector boson loop (generic name in \SARAH: \verb"FFV"):
\begin{eqnarray*}
\Sigma^S(p^2) &=& - 4 c_S c_C c_R m_F \Gamma^1_L \Gamma^{2,*}_R B_0(p^2,m_F^2,m_S^2) \\
\Sigma^R(p^2) &=& - c_S c_C c_R \Gamma^1_L \Gamma^{2,*}_L B_1(p^2,m_F^2,m_S^2) \\
\Sigma^L(p^2) &=& - c_S c_C c_R \Gamma^1_R \Gamma^{2,*}_R B_1(p^2,m_F^2,m_S^2) 
\end{eqnarray*}
\end{enumerate}

\paragraph{Corrections to scalar}
\begin{enumerate}
\item Fermion loop (generic name in \SARAH: \verb"FFS"): 
\begin{equation}
\label{eq:SE_FFS}
\Pi(p^2) = c_S c_C c_R \left((\Gamma^1_L \Gamma^{2,*}_L + \Gamma^1_R \Gamma^{2,*}_R) G_0(p^2,m_F^2,m_S^2) +  (\Gamma^1_L \Gamma^{2,*}_R + \Gamma^1_R \Gamma^{2,*}_L) B_0(p^2,m_F^2,m_S^2) \right)
\end{equation}
\item Scalar loop (two 3-point interactions, generic name in \SARAH: \verb"SSS"):
\begin{equation}
\Pi(p^2) = c_S c_C c_R \Gamma^1 \Gamma^{2,*} B_0(p^2,m_F^2,m_S^2) 
\end{equation}
\item Scalar loop (4-point interaction, generic name in \SARAH: \verb"SSSS"):
\begin{equation}
\label{eq:SE_SSSS}
\Pi(p^2) =  - c_S c_C \Gamma A_0(m_S^2) 
\end{equation}
\item Vector boson-scalar loop (generic name in \SARAH: \verb"SSV"):
\begin{equation}
\label{eq:SE_SSV}
\Pi(p^2) = c_S c_C c_R \Gamma^1 \Gamma^{2,*} F_0(p^2,m_F^2,m_S^2) 
\end{equation}
\item Vector boson loop (two 3-point interactions, generic name in \SARAH: \verb"SVV"):
\begin{equation}
\Pi(p^2) =  c_S c_C c_R \frac{7}{2} \Gamma^1 \Gamma^{2,*} B_0(p^2,m_F^2,m_S^2) 
\end{equation}
\item Vector boson loop (4-point interaction, generic name in \SARAH: \verb"SSVV"):
\begin{equation}
\Pi(p^2) =   c_S c_C \Gamma A_0(m_V^2) 
\end{equation}
\end{enumerate}

\paragraph{Corrections to vector boson}
\begin{enumerate}
\item Fermion loop (generic name in \SARAH: \verb"FFV"):
\begin{equation}
\Pi^T(p^2) = c_S c_C c_R \left((|\Gamma^1_L|^2+|\Gamma^1_R|^2) H_0(p^2,m_V^2,m_F^2)+ 4 \Re(\Gamma^1_L \Gamma^2_R)B_0(p^2,m_V^2,m_F^2) \right)
\end{equation}
\item Scalar loop (generic name in \SARAH: \verb"SSV"):
\begin{equation}
\Pi^T(p^2) = -4 c_S c_C c_R |\Gamma|^2 B_{22}(p^2,m_{S_1}^2,m_{S_2}^2)
\end{equation}
\item Vector boson loop (generic name in \SARAH: \verb"VVV"):
\begin{equation}
\Pi^T(p^2) =  |\Gamma|^2  c_S c_C c_R \left(-(4 p^2 + m_{V_1}^2 + m_{V_2}^2 ) B_0(p^2,m_{V_1}^2,m_{V_1}^2) - 8 B_{22}(p^2,m_{S_1}^2,m_{S_2}^2) \right)
\end{equation}
\item Vector-scalar loop (generic name in \SARAH: \verb"SVV"):
\begin{equation}
\Pi^T(p^2) =  |\Gamma|^2 c_S c_C c_R B_0(p^2,m_V^2,m_S^2)
\end{equation}
\end{enumerate}
We need here only the diagrams involving three-point interactions because the four-point interactions are related to them due to gauge invariance. 
\subsubsection{One-loop corrections to masses} 
\label{sec:OneLoopMass}
The one-loop self-energies can be used to calculate the one-loop masses and mass matrices. For the one-loop corrections of scalars, the radiatively corrected mass matrix is 
\begin{equation}
	m^{2,S}_{1L}(p^2_i) = m^{2,S}_{T} - \Pi_{S S}(p^2_i) , 
\end{equation}
while the one-loop mass of a vector boson \(V\) is given by
\begin{equation}
 m^{2,V}_{1L}(Q) =  m^{2,V}_T + \mathrm{Re}\big\{ \Pi^T_{VV}(m^{2,V}_T) \big\}.  
\end{equation}
According to the conventions of the counter terms of \cite{Pierce:1996zz}, the one-loop mass matrices \(M^{\tilde\chi^0}_{1L}\) of Majorana fermions are connected to the one-loop self-energies and tree-level mass matrix \(M^{\tilde\chi^0}_T\)  by
\begin{eqnarray}
M^{\tilde\chi^0}_{1L} (p^2_i) &=& M^{\tilde\chi^0}_T - 
\frac{1}{2} \bigg[ \Sigma^0_S(p^2_i) + \Sigma^{0,T}_S(p^2_i)
 + \left(\Sigma^{0,T}_L(p^2_i)+   \Sigma^0_R(p^2_i)\right) M^{\tilde\chi^0}_T
 \nonumber \\
&& \hspace{16mm}
+ M^{\tilde\chi^0}_T \left(\Sigma^{0,T}_R(p^2_i) +  \Sigma^0_L(p^2_i) \right) \bigg] .
\end{eqnarray}
In the case of Dirac fermions, the one-loop corrected mass matrix is 
\begin{eqnarray}
M^{\tilde\chi^+}_{1L}(p^2_i) =  M^{\tilde\chi^+}_T - \Sigma^+_S(p^2_i)
 - \Sigma^+_R(p^2_i) M^{\tilde \chi^+}_T - M^{\tilde \chi^+}_T \Sigma^+_L(p^2_i) .
\end{eqnarray}

\section{The minimal supersymmetric standard model}
\label{sec:MSSM}
\subsection{Vector superfields} 
\begin{center} 
\begin{tabular}{|c|c|c|c|c|c|} 
\hline \hline 
SF & Spin \(\frac{1}{2}\) & Spin 1 & \(SU(N)\) & Coupling & Name \\ 
 \hline 
\(\hat{B}\) & \(\lambda_{\tilde{B}}\) & \(B\) & \(U(1)\) & \(g_1\) &\text{hypercharge}\\ 
\(\hat{W}\) & \(\lambda_{\tilde{W}}\) & \(W\) & \(\text{SU}(2)\) & \(g_2\) &\text{left}\\ 
\(\hat{g}\) & \(\lambda_{\tilde{g}}\) & \(g\) & \(\text{SU}(3)\) & \(g_3\) &\text{color}\\ 
\hline \hline
\end{tabular} 
\end{center} 
\subsection{Chiral superfields} 
\begin{center} 
\begin{tabular}{|c|c|c|c|c|c|} 
\hline \hline 
SF & Spin 0 & Spin \(\frac{1}{2}\) & Generations & \((U(1)\otimes\, \text{SU}(2)\otimes\, \text{SU}(3))\) \\ 
\hline 
\(\hat{q}\) & \(\tilde{q}\) & \(q\) & 3 & \((\frac{1}{6},{\bf 2},{\bf 3}) \) \\ 
\(\hat{l}\) & \(\tilde{l}\) & \(l\) & 3 & \((-\frac{1}{2},{\bf 2},{\bf 1}) \) \\ 
\(\hat{H}_d\) & \(H_d\) & \(\tilde{H}_d\) & 1 & \((-\frac{1}{2},{\bf 2},{\bf 1}) \) \\ 
\(\hat{H}_u\) & \(H_u\) & \(\tilde{H}_u\) & 1 & \((\frac{1}{2},{\bf 2},{\bf 1}) \) \\ 
\(\hat{d}\) & \(\tilde{d}_R^*\) & \(d_R^*\) & 3 & \((\frac{1}{3},{\bf 1},{\bf \overline{3}}) \) \\ 
\(\hat{u}\) & \(\tilde{u}_R^*\) & \(u_R^*\) & 3 & \((-\frac{2}{3},{\bf 1},{\bf \overline{3}}) \) \\ 
\(\hat{e}\) & \(\tilde{e}_R^*\) & \(e_R^*\) & 3 & \((1,{\bf 1},{\bf 1}) \) \\ 
\hline \hline
\end{tabular} 
\end{center} 
\subsection{Superpotential and Lagrangian} 
\paragraph{Superpotential} 
\begin{align} 
W = & \,  Y_u\,\hat{u}\,\hat{q}\,\hat{H}_u\,- Y_d \,\hat{d}\,\hat{q}\,\hat{H}_d\,- Y_e \,\hat{e}\,\hat{l}\,\hat{H}_d\,+\mu\,\hat{H}_u\,\hat{H}_d\,\end{align} 
\paragraph{Soft-breaking terms} 
\begin{align} 
- L_{SB,W} = \, & - H_d^0 H_u^0 B_{\mu} +H_d^- H_u^+ B_{\mu} +H_d^0 \tilde{d}^*_{R,{i \alpha}} \delta_{\alpha\beta} \tilde{d}_{L,{j \beta}} T_{d,{i j}} - H_d^- \tilde{d}^*_{R,{i \alpha}} \delta_{\alpha\beta} \tilde{u}_{L,{j \beta}} T_{d,{i j}} \nonumber \\ 
 &+H_d^0 \tilde{e}^*_{R,{i}} \tilde{e}_{L,{j}} T_{e,{i j}} - H_d^- \tilde{e}^*_{R,{i}} \tilde{\nu}_{L,{j}} T_{e,{i j}} - H_u^+ \tilde{u}^*_{R,{i \alpha}} \delta_{\alpha\beta} \tilde{d}_{L,{j \beta}} T_{u,{i j}} +H_u^0 \tilde{u}^*_{R,{i \alpha}} \delta_{\alpha\beta} \tilde{u}_{L,{j \beta}} T_{u,{i j}} + \mbox{h.c.} \\ 
- L_{SB,\phi} = \, & +m_{H_d}^2 |H_d^0|^2 +m_{H_d}^2 |H_d^-|^2 +m_{H_u}^2 |H_u^0|^2 +m_{H_u}^2 |H_u^+|^2 +\tilde{d}^*_{L,{j \beta}} \delta_{\alpha\beta} m_{q,{i j}}^{2} \tilde{d}_{L,{i \alpha}} \nonumber \\ 
 &+\tilde{d}^*_{R,{i \alpha}} \delta_{\alpha\beta} m_{d,{i j}}^{2} \tilde{d}_{R,{j \beta}} +\tilde{e}^*_{L,{j}} m_{l,{i j}}^{2} \tilde{e}_{L,{i}} +\tilde{e}^*_{R,{i}} m_{e,{i j}}^{2} \tilde{e}_{R,{j}} +\tilde{u}^*_{L,{j \beta}} \delta_{\alpha\beta} m_{q,{i j}}^{2} \tilde{u}_{L,{i \alpha}} \nonumber \\ 
 &+\tilde{u}^*_{R,{i \alpha}} \delta_{\alpha\beta} m_{u,{i j}}^{2} \tilde{u}_{R,{j \beta}} +\tilde{\nu}^*_{L,{j}} m_{l,{i j}}^{2} \tilde{\nu}_{L,{i}} \\ 
- L_{SB,\lambda} = \, & \frac{1}{2} \left( \lambda_{\tilde{B}}^{2} M_1  + M_2 \lambda_{{\tilde{W}},{i}}^{2}  + M_3 \lambda_{{\tilde{g}},{i}}^{2} + \mbox{h.c.} \right) 
\end{align} 
\subsubsection{Gauge fixing terms} 
\paragraph{Gauge fixing terms for gauge eigenstates } 
\begin{align} 
L_{GF} = \, &-\frac{1}{2 \xi_{G}} \partial_{\mu}g_\alpha   -\frac{1}{2 \xi_{W}} \partial_{\mu}W^i 
\end{align} 
\paragraph{Gauge fixing terms for mass eigenstates after EWSB } 
\begin{align} 
L_{GF} = \, &-\frac{1}{2 \xi_{P}} \partial_{\mu}\gamma  -\frac{1}{2 \xi_{G}} \partial_{\mu}g_\alpha   -\frac{1}{2 \xi_{Z}}  \Big(- A^0_{{1}} m_{Z} \xi_{Z}  + \partial_{\mu}Z\Big) - \frac{1}{\xi_{W}} \Big(i H^-_{{1}} m_{W^-} \xi_{W}  + \partial_{\mu}W^-\Big)
\end{align} 
\subsection{Vacuum expectation values}
\label{sec:VEVsMSSM}
\begin{align} 
H_d^0 =  \, \frac{1}{\sqrt{2}} \left( \phi_{d}  + i \sigma_{d}  +  v_d  \right) \, , \hspace{1cm} 
H_u^0 =  \, \frac{1}{\sqrt{2}} \left( \phi_{u}    + i  \sigma_{u}  +  v_u \right) 
\end{align}

\subsection{Rotations of vector bosons and gauginos after EWSB} 
\label{sec:RotGaugeMSSM}
\begin{align} 
W^-_{{1 \rho}} = & \,\frac{1}{\sqrt{2}} W^-_{{\rho}}  + \frac{1}{\sqrt{2}} W^+_{{\rho}} \, , \hspace{1cm} 
W^-_{{2 \rho}} =  \,-i \frac{1}{\sqrt{2}} W^-_{{\rho}}  + i \frac{1}{\sqrt{2}} W^+_{{\rho}} \\ 
W^-_{{3 \rho}} = & \,\cos\Theta_W  Z_{{\rho}}  + \sin\Theta_W  \gamma_{{\rho}} \, , \hspace{1cm}  
B_{{\rho}} =  \,\cos\Theta_W  \gamma_{{\rho}}  - \sin\Theta_W  Z_{{\rho}} \\ 
\lambda_{{\tilde{W}},{1}} = & \,\frac{1}{\sqrt{2}} \tilde{W}^-  + \frac{1}{\sqrt{2}} \tilde{W}^+ \, , \hspace{1cm}  
\lambda_{{\tilde{W}},{2}} =  \,-i \frac{1}{\sqrt{2}} \tilde{W}^-  + i \frac{1}{\sqrt{2}} \tilde{W}^+ \, , \hspace{1cm}  
\lambda_{{\tilde{W}},{3}} =  \,\tilde{W}^0
\end{align} 

\subsection{Rotations in matter sector to mass eigenstates after EWSB}
\label{sec:RotMatterMSSM}
In the following, Greek letters \(\alpha_i,\beta_i\) refer to color indices and \(o_i, p_i\) to generations indices.  
\begin{enumerate} 
\item {\bf Mass matrix for neutralinos}, basis: \( \left(\lambda_{\tilde{B}}, \tilde{W}^0, \tilde{H}_d^0, \tilde{H}_u^0\right)\)
\begin{equation} 
m_{\tilde{\chi}^0} = \left( 
\begin{array}{cccc}
M_1 &0 &-\frac{1}{2} g_1 v_d  &\frac{1}{2} g_1 v_u \\ 
0 &M_2 &\frac{1}{2} g_2 v_d  &-\frac{1}{2} g_2 v_u \\ 
-\frac{1}{2} g_1 v_d  &\frac{1}{2} g_2 v_d  &0 &- \mu \\ 
\frac{1}{2} g_1 v_u  &-\frac{1}{2} g_2 v_u  &- \mu  &0\end{array} 
\right) 
\end{equation} 
This matrix is diagonalized by \(N\): 
\begin{equation} 
N m_{\tilde{\chi}^0} N^{\dagger} = m^{dia}_{\tilde{\chi}^0} 
\end{equation} 
with 
\begin{align} 
\lambda_{\tilde{B}} = \sum_{t_2}N^*_{j 1}\lambda^0_{{j}}\,, \hspace{1cm} 
\tilde{W}^0 = \sum_{t_2}N^*_{j 2}\lambda^0_{{j}} \,, \hspace{1cm}  
\tilde{H}_d^0 = \sum_{t_2}N^*_{j 3}\lambda^0_{{j}}\,, \hspace{1cm}  
\tilde{H}_u^0 = \sum_{t_2}N^*_{j 4}\lambda^0_{{j}}
\end{align} 
\item {\bf Mass matrix for charginos}, basis: \( \left(\tilde{W}^-, \tilde{H}_d^-\right)/\left(\tilde{W}^+, \tilde{H}_u^+\right) \) 
 
\begin{equation} 
m_{\tilde{\chi}^-} = \left( 
\begin{array}{cc}
M_2 &\frac{1}{\sqrt{2}} g_2 v_u \\ 
\frac{1}{\sqrt{2}} g_2 v_d  &\mu\end{array} 
\right) 
\end{equation} 
This matrix is diagonalized by \(U\) and \(V\) 
\begin{equation} 
U^* m_{\tilde{\chi}^-} V^{\dagger} = m^{dia}_{\tilde{\chi}^-} 
\end{equation} 
with 
\begin{align} 
\tilde{W}^- = \sum_{t_2}U^*_{j 1}\lambda^-_{{j}}\,, \hspace{1cm} 
\tilde{H}_d^- = \sum_{t_2}U^*_{j 2}\lambda^-_{{j}} \,, \hspace{1cm}  
\tilde{W}^+ = \sum_{t_2}V^*_{1 j}\lambda^+_{{j}}\,, \hspace{1cm} 
\tilde{H}_u^+ = \sum_{t_2}V^*_{2 j}\lambda^+_{{j}}
\end{align} 
\item {\bf Mass matrix for leptons}, basis: \( \left(e_{L,{{o_1}}}\right)/\left(e^*_{R,{{p_1}}}\right) \) 
 
\begin{equation} 
m_{e} = \left( 
\begin{array}{c}
\frac{1}{\sqrt{2}} v_d Y_{e,{{p_1} {o_1}}} \end{array} 
\right) 
\end{equation} 
This matrix is diagonalized by \(U^e_L\) and \(U^e_R\) 
\begin{equation} 
U^{e,*}_L m_{e} U_{R}^{e,\dagger} = m^{dia}_{e} 
\end{equation} 
with 
\begin{align} 
e_{L,{i}} = \sum_{t_2}U^{e,*}_{L,{j i}}E_{L,{j}}\,, \hspace{1cm} 
e_{R,{i}} = \sum_{t_2}U_{R,{i j}}^{e}E^*_{R,{j}}
\end{align} 
\item {\bf Mass matrix for down-quarks}, basis: \( \left(d_{L,{{o_1} {\alpha_1}}}\right)/\left(d^*_{R,{{p_1} {\beta_1}}}\right) \) 
 
\begin{equation} 
m_{d} = \left( 
\begin{array}{c}
\frac{1}{\sqrt{2}} v_d \delta_{{\alpha_1}{\beta_1}} Y_{d,{{p_1} {o_1}}} \end{array} 
\right) 
\end{equation} 
This matrix is diagonalized by \(U^d_L\) and \(U^d_R\) 
\begin{equation} 
U^{d,*}_L m_{d} U_{R}^{d,\dagger} = m^{dia}_{d} 
\end{equation} 
with 
\begin{align} 
d_{L,{i \alpha}} = \sum_{t_2}U^{d,*}_{L,{j i}}D_{L,{j \alpha}}\,, \hspace{1cm}  
d_{R,{i \alpha}} = \sum_{t_2}U_{R,{i j}}^{d}D^*_{R,{j \alpha}}
\end{align} 
\item {\bf Mass matrix for up-quarks}, basis: \( \left(u_{L,{{o_1} {\alpha_1}}}\right)/\left(u^*_{R,{{p_1} {\beta_1}}}\right) \) 
 
\begin{equation} 
m_{u} = \left( 
\begin{array}{c}
\frac{1}{\sqrt{2}} v_u \delta_{{\alpha_1}{\beta_1}} Y_{u,{{p_1} {o_1}}} \end{array} 
\right) 
\end{equation} 
This matrix is diagonalized by \(U^u_L\) and \(U^u_R\) 
\begin{equation} 
U^{u,*}_L m_{u} U_{R}^{u,\dagger} = m^{dia}_{u} 
\end{equation} 
with 
\begin{align} 
u_{L,{i \alpha}} = \sum_{t_2}U^{u,*}_{L,{j i}}U_{L,{j \alpha}}\,, \hspace{1cm}  
u_{R,{i \alpha}} = \sum_{t_2}U_{R,{i j}}^{u}U^*_{R,{j \alpha}}
\end{align} 
\item {\bf Mass matrix for down-squarks}, basis: \( \left(\tilde{d}_{L,{{o_1} {\alpha_1}}}/\tilde{d}_{R,{{o_2} {\alpha_2}}}\right), \left(\tilde{d}^*_{L,{{p_1} {\beta_1}}}, \tilde{d}^*_{R,{{p_2} {\beta_2}}}\right) \) 
\begin{align} 
m_{11} &= \frac{1}{24} \delta_{{\alpha_1}{\beta_1}} \Big(12 \Big(2 m_{q,{{o_1} {p_1}}}^{2}  + v_{d}^{2} \sum_{a=1}^{3}Y^*_{d,{a {p_1}}} Y_{d,{a {o_1}}}  \Big) - \Big(3 g_{2}^{2}  + g_{1}^{2}\Big)\Big(- v_{u}^{2}  + v_{d}^{2}\Big)\delta_{{o_1}{p_1}} \Big)\\
m_{12} &=  \frac{1}{\sqrt{2}} \delta_{{\alpha_1}{\beta_2}} \Big(v_d T_{d,{{p_2} {o_1}}}  - v_u \mu^* Y_{d,{{p_2} {o_1}}} \Big) \\
m_{22} &= \frac{1}{12} \delta_{{\alpha_2}{\beta_2}} \Big(6 \Big(2 m_{d,{{p_2} {o_2}}}^{2}  + v_{d}^{2} \sum_{a=1}^{3}Y^*_{d,{{o_2} a}} Y_{d,{{p_2} a}}  \Big) + g_{1}^{2} \Big(- v_{d}^{2}  + v_{u}^{2}\Big)\delta_{{o_2}{p_2}} \Big)
\end{align} 
This matrix is diagonalized by \(Z^D\): 
\begin{equation} 
Z^D m^2_{\tilde{d}} Z^{D,\dagger} = m^{dia}_{2,\tilde{d}} 
\end{equation} 
with 
\begin{align} 
\tilde{d}_{L,{i \alpha}} = \sum_{t_2}Z^{D,*}_{j i}\tilde{d}_{{j \alpha}}\,, \hspace{1cm} 
\tilde{d}_{R,{i \alpha}} = \sum_{t_2}Z^{D,*}_{j i}\tilde{d}_{{j \alpha}}
\end{align} 
\item {\bf Mass matrix for sneutrinos}, basis: \( \left(\tilde{\nu}_{L,{{o_1}}}\right)/ \left(\tilde{\nu}^*_{L,{{p_1}}}\right) \) 
 
\begin{equation} 
m^2_{\tilde{\nu}} = \left( 
\begin{array}{c}
\frac{1}{8} \Big(8 m_{l,{{o_1} {p_1}}}^{2}  + \Big(g_{1}^{2} + g_{2}^{2}\Big)\Big(- v_{u}^{2}  + v_{d}^{2}\Big)\delta_{{o_1}{p_1}} \Big)\end{array} 
\right) 
\end{equation} 
This matrix is diagonalized by \(Z^V\): 
\begin{equation} 
Z^V m^2_{\tilde{\nu}} Z^{V,\dagger} = m^{dia}_{2,\tilde{\nu}} 
\end{equation} 
with 
\begin{align} 
\tilde{\nu}_{L,{i}} = \sum_{t_2}Z^{V,*}_{j i}\tilde{\nu}_{{j}}
\end{align} 
\item {\bf Mass matrix for up-squarks}, basis: \( \left(\tilde{u}_{L,{{o_1} {\alpha_1}}}, \tilde{u}_{R,{{o_2} {\alpha_2}}}\right)/ \left(\tilde{u}^*_{L,{{p_1} {\beta_1}}}, \tilde{u}^*_{R,{{p_2} {\beta_2}}}\right) \) 
\begin{align} 
m_{11} &= \frac{1}{24} \delta_{{\alpha_1}{\beta_1}} \Big(12 \Big(2 m_{q,{{o_1} {p_1}}}^{2}  + v_{u}^{2} \sum_{a=1}^{3}Y^*_{u,{a {p_1}}} Y_{u,{a {o_1}}}  \Big) - \Big(-3 g_{2}^{2}  + g_{1}^{2}\Big)\Big(- v_{u}^{2}  + v_{d}^{2}\Big)\delta_{{o_1}{p_1}} \Big)\\ 
m_{12} &= \frac{1}{\sqrt{2}} \delta_{{\alpha_1}{\beta_2}} \Big(- v_d \mu^* Y_{u,{{p_2} {o_1}}}  + v_u T_{u,{{p_2} {o_1}}} \Big) \\
m_{22} &= \frac{1}{6} \delta_{{\alpha_2}{\beta_2}} \Big(3 v_{u}^{2} \sum_{a=1}^{3}Y^*_{u,{{o_2} a}} Y_{u,{{p_2} a}}   + 6 m_{u,{{p_2} {o_2}}}^{2}  + g_{1}^{2} \Big(- v_{u}^{2}  + v_{d}^{2}\Big)\delta_{{o_2}{p_2}} \Big)
\end{align} 
This matrix is diagonalized by \(Z^U\): 
\begin{equation} 
Z^U m^2_{\tilde{u}} Z^{U,\dagger} = m^{dia}_{2,\tilde{u}} 
\end{equation} 
with 
\begin{align} 
\tilde{u}_{L,{i \alpha}} = \sum_{t_2}Z^{U,*}_{j i}\tilde{u}_{{j \alpha}}\,, \hspace{1cm} 
\tilde{u}_{R,{i \alpha}} = \sum_{t_2}Z^{U,*}_{j i}\tilde{u}_{{j \alpha}}
\end{align} 
\item {\bf Mass matrix for sleptons}, basis: \( \left(\tilde{e}_{L,{{o_1}}}, \tilde{e}_{R,{{o_2}}}\right)/\left(\tilde{e}^*_{L,{{p_1}}}, \tilde{e}^*_{R,{{p_2}}}\right) \) 
\begin{align} 
m_{11} &= \frac{1}{8} \Big(4 v_{d}^{2} \sum_{a=1}^{3}Y^*_{e,{a {p_1}}} Y_{e,{a {o_1}}}   + 8 m_{l,{{o_1} {p_1}}}^{2}  + \Big(- g_{2}^{2}  + g_{1}^{2}\Big)\Big(- v_{u}^{2}  + v_{d}^{2}\Big)\delta_{{o_1}{p_1}} \Big)\\ 
m_{12} &= \frac{1}{\sqrt{2}} \Big(v_d T_{e,{{p_2} {o_1}}}  - v_u \mu^* Y_{e,{{p_2} {o_1}}} \Big) \\
m_{22} &= \frac{1}{4} \Big(2 v_{d}^{2} \sum_{a=1}^{3}Y^*_{e,{{o_2} a}} Y_{e,{{p_2} a}}   + 4 m_{e,{{p_2} {o_2}}}^{2}  + g_{1}^{2} \Big(- v_{d}^{2}  + v_{u}^{2}\Big)\delta_{{o_2}{p_2}} \Big)
\end{align} 
This matrix is diagonalized by \(Z^E\): 
\begin{equation} 
Z^E m^2_{\tilde{e}} Z^{E,\dagger} = m^{dia}_{2,\tilde{e}} 
\end{equation} 
with 
\begin{align} 
\tilde{e}_{L,{i}} = \sum_{t_2}Z^{E,*}_{j i}\tilde{e}_{{j}}\,, \hspace{1cm} 
\tilde{e}_{R,{i}} = \sum_{t_2}Z^{E,*}_{j i}\tilde{e}_{{j}}
\end{align} 
\item {\bf Mass matrix for scalar Higgs}, basis: \( \left(\phi_{d}, \phi_{u}\right)\)
\begin{equation} 
m^2_{h} = \left( 
\begin{array}{cc}
m_{H_d}^2  +  |\mu|^2  +\frac{1}{8} \Big(g_{1}^{2} + g_{2}^{2}\Big)\Big(3 v_{d}^{2}  - v_{u}^{2} \Big) & -{\Re\Big(B_{\mu}\Big)}  - \frac{1}{4}\Big(g_{1}^{2} + g_{2}^{2}\Big)v_d v_u \\ 
-{\Re\Big(B_{\mu}\Big)}  - \frac{1}{4}\Big(g_{1}^{2} + g_{2}^{2}\Big)v_d v_u &m_{H_u}^2  +  |\mu|^2  - \frac{1}{8}\Big(g_{1}^{2} + g_{2}^{2}\Big)\Big(-3 v_{u}^{2}  + v_{d}^{2}\Big)\end{array} 
\right) 
\end{equation} 
This matrix is diagonalized by \(Z^H\): 
\begin{equation} 
Z^H m^2_{h} Z^{H,\dagger} = m^{dia}_{2,h} 
\end{equation} 
with 
\begin{align} 
\phi_{d} = \sum_{t_2}Z_{{j 1}}^{H}h_{{j}}\,, \hspace{1cm} 
\phi_{u} = \sum_{t_2}Z_{{j 2}}^{H}h_{{j}}
\end{align} 
The mixing matrix can be parametrized by 
\begin{equation} 
Z^H= \, \left( 
\begin{array}{cc} 
- \sin\alpha   & \cos\alpha  \\ 
 \cos\alpha  & \sin\alpha \end{array} 
\right) 
\end{equation} 
\item {\bf Mass matrix for pseudo scalar Higgs}, basis: \( \left(\sigma_{d}, \sigma_{u}\right)\) 
\begin{equation} 
m^2_{A^0} = \left( 
\begin{array}{cc}
m_{H_d}^2  +  |\mu|^2  + \frac{1}{8}  \Big(g_{1}^{2} + g_{2}^{2}\Big)\Big(- v_{u}^{2}  + v_{d}^{2}\Big) &{\Re\Big(B_{\mu}\Big)}\\ 
{\Re\Big(B_{\mu}\Big)} & m_{H_u}^2  +  |\mu|^2  -\frac{1}{8}  \Big(g_{1}^{2} + g_{2}^{2}\Big)\Big(- v_{u}^{2}  + v_{d}^{2}\Big)\end{array} 
\right) 
\end{equation} 
This matrix is diagonalized by \(Z^A\): 
\begin{equation} 
Z^A m^2_{A^0} Z^{A,\dagger} = m^{dia}_{2,A^0} 
\end{equation} 
with 
\begin{align} 
\sigma_{d} = \sum_{t_2}Z_{{j 1}}^{A}A^0_{{j}}\,, \hspace{1cm} 
\sigma_{u} = \sum_{t_2}Z_{{j 2}}^{A}A^0_{{j}}
\end{align} 
The mixing matrix can be parametrized by 
\begin{equation} 
Z^A= \, \left( 
\begin{array}{cc} 
- \cos\beta   & \sin\beta  \\ 
 \sin\beta  & \cos\beta \end{array} 
\right) 
\end{equation} 
\item {\bf Mass matrix for charged Higgs}, basis: \( \left(H_d^-, H_u^{+,*}\right)\) 
\begin{equation} 
m^2_{H^-} = \left( 
\begin{array}{cc}
 m_{H_d}^2  +  |\mu|^2  +  \frac{1}{8} \Big(g_{1}^{2} + g_2^2\Big)\Big(v_{d}^{2}  -  v_{u}^{2}\Big) &\frac{1}{4} g_{2}^{2} v_d v_u  + B_{\mu}\\ 
\frac{1}{4} g_{2}^{2} v_d v_u  + B_{\mu}^* & m_{H_u}^2  + |\mu|^2  + \frac{1}{8} \Big(g_2^2 - g_{1}^{2}\Big)\Big(v_{d}^{2}  -  v_{u}^{2}\Big)\end{array} 
\right) 
\end{equation} 
This matrix is diagonalized by \(Z^+\): 
\begin{equation} 
Z^+ m^2_{H^-} Z^{+,\dagger} = m^{dia}_{2,H^-} 
\end{equation} 
with 
\begin{align} 
H_d^- = \sum_{t_2}Z^{+,*}_{j 1}H^-_{{j}}\,, \hspace{1cm} 
H_u^+ = \sum_{t_2}Z_{{j 2}}^{+}H^+_{{j}}
\end{align} 
The mixing matrix can be parametrized by 
\begin{equation} 
Z^+= \, \left( 
\begin{array}{cc} 
- \cos\beta   & \sin\beta  \\ 
 \sin\beta  & \cos\beta \end{array} 
\right) 
\end{equation} 
\end{enumerate} 
\subsection{Tadpole equations}
\begin{align} 
\frac{\partial V}{\partial v_d} &= \frac{1}{8} \Big(8 v_d |\mu|^2  -8 v_u {\Re\Big(B_{\mu}\Big)}  + v_d \Big(8 m_{H_d}^2  + g_{1}^{2} v_{d}^{2}  - g_{1}^{2} v_{u}^{2}  + g_{2}^{2} v_{d}^{2}  - g_{2}^{2} v_{u}^{2} \Big)\Big)\\ 
\frac{\partial V}{\partial v_u} &= \frac{1}{8} \Big(-8 v_d {\Re\Big(B_{\mu}\Big)}  + 8 v_u |\mu|^2  + v_u \Big(8 m_{H_u}^2  - g_{1}^{2} v_{d}^{2}  + g_{1}^{2} v_{u}^{2}  - g_{2}^{2} v_{d}^{2}  + g_{2}^{2} v_{u}^{2} \Big)\Big)
\end{align} 
\section{Particles and parameters of the MSSM in \SARAH}
\label{sec:SARAH_MSSM}
\subsection*{Particles}
We show here only the eigenstates after EWSB
\begin{enumerate} 
\item Fermions \\ 
\begin{center} 
\begin{tabular}{|cc|cc|} 
\hline 
\(\tilde{\chi}^-_{{i}} \)  & \( \verb"Cha[{generation}]" \) & 
\(\tilde{\chi}^0_{{i}} \) & \( \verb"Chi[{generation}]" \) \\ 
 \(d_{{i \alpha}} \) & \( \verb"Fd[{generation, color}]"  \)  & 
 \(e_{{i}} \) & \( \verb"Fe[{generation}]" \)  \\ 
 \(u_{{i \alpha}} \) & \( \verb"Fu[{generation, color}]" \)  & 
 \(\nu_{{i}} \) & \( \verb"Fv[{generation}]" \)  \\ 
 \(\tilde{g}_{{i}} \) & \( \verb"Glu[{generation}]" \)  & {} & {} \\
 \hline 
\end{tabular} 
\end{center} 
\item Scalars \\ 
\begin{center} 
\begin{tabular}{|cc|cc|} 
\hline 
\(\tilde{d}_{{i \alpha}}\) & \verb"Sd[{generation, color}]"  & \(\tilde{\nu}_{{i}}\) & \verb"Sv[{generation}]" \\ 
\(\tilde{u}_{{i \alpha}}\) & \verb"Su[{generation, color}]"  & \(\tilde{e}_{{i}}\) & \verb"Se[{generation}]" \\ 
\(h_{{i}}\) & \verb"hh[{generation}]"  & \(A^0_{{i}}\) & \verb"Ah[{generation}]" \\ 
\(H^-_{{i}}\) & \verb"Hpm[{generation}]" & & \\ 
 \hline 
\end{tabular} 
\end{center} 
\item Vector bosons \\ 
\begin{center} 
\begin{tabular}{|cc|cc|} 
\hline 
\(g_{{i \rho}}\) & \verb"VG[{generation, lorentz}]"  & \(W^-_{{\rho}}\) & \verb"VWm[{lorentz}]" \\ 
\(\gamma_{{\rho}}\) & \verb"VP[{lorentz}]"  & \(Z_{{\rho}}\) & \verb"VZ[{lorentz}]" \\ 
\hline 
\end{tabular} 
\end{center} 
\item Ghosts \\ 
\begin{center} 
\begin{tabular}{|cc|cc|} 
\hline 
\(\eta^G_{{i}}\) & \verb"gG[{generation}]"  & \(\eta^-\) & \verb"gWm" \\ 
\(\eta^+\) & \verb"gWmC"  & \(\eta^{\gamma}\) & \verb"gP" \\ 
\(\eta^Z\) & \verb"gZ" & & \\ 
 \hline 
\end{tabular} 
\end{center} 
\end{enumerate} 

\subsection*{Parameters} 
\begin{center} 
\begin{tabular}{|cc|cc|cc|} 
\hline 
\(g_1\) & \verb"g1"  & \(g_2\) & \verb"g2"  & \(g_3\) & \verb"g3" \\ 
\(Y_u\) & \verb"Yu"  & \(T_u\) & \verb"T[Yu]"  & \(Y_d\) & \verb"Yd" \\ 
\(T_d\) & \verb"T[Yd]"  & \(Y_e\) & \verb"Ye"  & \(T_e\) & \verb"T[Ye]" \\ 
\(\mu\) & \verb"\[Mu]"  & \(B_{\mu}\) & \verb"B[\[Mu]]"  & \(m_q^2\) & \verb"mq2" \\ 
\(m_l^2\) & \verb"ml2"  & \(m_{H_d}^2\) & \verb"mHd2"  & \(m_{H_u}^2\) & \verb"mHu2" \\ 
\(m_d^2\) & \verb"md2"  & \(m_u^2\) & \verb"mu2"  & \(m_e^2\) & \verb"me2" \\ 
\(M_1\) & \verb"MassB"  & \(M_2\) & \verb"MassWB"  & \(M_3\) & \verb"MassG" \\ 
\(v_d\) & \verb"vd"  & \(v_u\) & \verb"vu"  & \(\Theta_W\) & \verb"ThetaW" \\ 
\(\phi_{\tilde{g}}\) & \verb"PhaseGlu"  & \(Z^D\) & \verb"ZD"  & \(Z^V\) & \verb"ZV" \\ 
\(Z^U\) & \verb"ZU"  & \(Z^E\) & \verb"ZE"  & \(Z^H\) & \verb"ZH" \\ 
\(Z^A\) & \verb"ZA"  & \(Z^+\) & \verb"ZP"  & \(N\) & \verb"ZN" \\ 
\(U\) & \verb"UM"  & \(V\) & \verb"UP"  & \(U^e_L\) & \verb"ZEL" \\ 
\(U^e_R\) & \verb"ZER"  & \(U^d_L\) & \verb"ZDL"  & \(U^d_R\) & \verb"ZDR" \\ 
\(U^u_L\) & \verb"ZUL"  & \(U^u_R\) & \verb"ZUR"  & \(\alpha\) & \verb"\[Alpha]" \\ 
\(\beta\) & \verb"\[Beta]" & && & \\ 
 \hline 
\end{tabular} 
\end{center} 
\section{Changes in comparison to version 1 of \SARAH}
\label{app:changes}
We want shortly give an overview to the user about the main changes in the new version of \SARAH. 
\begin{enumerate}
 \item New physical output:
   \begin{enumerate}
    \item One- and two-loop renormalization group equations.
    \item One-loop self-energies and one-loop corrected tadpoles.
    \item All irreducible representations of chiral superfields possible .
    \item Some representation theory with regard to \(SU(N)\) gauge groups.
    \item Implicit charge indices are no longer restricted to \(SU(2)_L\).
    \item Check for charge conservation of a model.
   \end{enumerate}
 \item Changes in definition of models:
    \begin{enumerate}
      \item {\tt DEFINITION} statements to structure the model file.
      \item Definition of global properties of parameters and particles.
      \item Additional interactions added to the Lagrangian also rotated to new basis.
      \item Possibility to add phases to particles.
      \item Possibility to decompose one field with several generations in flavor eigenstates.
      \item Soft-breaking terms are named to SLHA 2 conventions.
      \item Improved routines to read LesHouches files.
    \end{enumerate}
 \item Changes in output
   \begin{enumerate}
      \item Speed of \CalcHep/\CompHep and \LaTeX output significantly improved.
      \item Adding of running coupling in \CalcHep model file.
      \item Possible suppression of splitting of four-scalar interactions in \CalcHep.
      \item Typesetting in \LaTeX{} output improved.
      \item Overview of superfields, important terms of the Lagrangian, mass matrices, tadpole equations, RGEs and self-energies also added to the \LaTeX output.
      \item New commands {\tt MassMatrix} and {\tt TadpoleEquation} to have easier access to information.
   \end{enumerate}
 \item New models
   \begin{enumerate}
      \item MSSM with trilinear R-parity violation.
      \item Singlet Extended Minimal Supersymmetric Standard Model (MSSM).
      \item The $U(1)$-Extended Minimal Supersymmetric Standard Model (UMSSM).
      \item The  Secluded $U(1)$-Extended Minimal Supersymmetric Standard Model (sMSSM).
      \item The near-to-Minimal Supersymmetric Standard Model (nMSSM).
   \end{enumerate}
\end{enumerate}

\end{appendix}

\end{document}